\documentclass[acmsmall, screen]{acmart}

\usepackage[normalem]{ulem}
\usepackage{bm}

\newtheorem{definition}{Definition}

\usepackage{subfigure} 
\usepackage{graphicx}
\usepackage{ragged2e}
\usepackage[justification=centering]{caption}

\usepackage{amsmath}
\usepackage{booktabs}
\usepackage{multirow}

\usepackage{tikz}
\usetikzlibrary{snakes}
\usepackage{forest}
\usepackage{xcolor}

\newcommand{\finding}[2]{
    \begin{center}
        \fcolorbox{black}{gray!10}{\parbox{0.98\textwidth}{
            \textbf{Summary for RQ{#1}:}{#2}
        }
    }
    \end{center}
}

\AtBeginDocument{%
  \providecommand\BibTeX{{%
    \normalfont B\kern-0.5em{\scshape i\kern-0.25em b}\kern-0.8em\TeX}}}

\setcopyright{acmcopyright}
\copyrightyear{2023}
\acmYear{2023}
\acmDOI{10.1145/1122445.1122456}

\acmJournal{TOSEM}
\acmVolume{0}
\acmNumber{0}
\acmArticle{1}
\acmMonth{0}

\begin{document}

\title{A Survey of Source Code Search: A 3-Dimensional Perspective}

\author{Weisong Sun}
\email{weisongsun@smail.nju.edu.cn}
\email{weisong.sun@ntu.edu.sg}
\orcid{0000-0001-9236-8264}
\affiliation{
  \institution{State Key Laboratory for Novel Software Technology, Nanjing University}
  \city{Nanjing}
  \state{Jiangsu}
  \country{China}
  \postcode{210093}
}
\affiliation{
  \institution{School of Computer Science and Engineering, Nanyang Technological University}
  \state{50 Nanyang Avenue}
  \country{Singapore}
  \postcode{639798}
}

\author{Chunrong Fang} \email{fangchunrong@nju.edu.cn}
\orcid{0000-0002-9930-7111}
\authornote{\textbf{Chunrong Fang is the corresponding author.}}
\affiliation{
  \institution{State Key Laboratory for Novel Software Technology, Nanjing University}
  \city{Nanjing}
  \state{Jiangsu}
  \country{China}
  \postcode{210093}
}

\author{Yifei Ge}
\email{gyf991213@126.com}
\orcid{0009-0009-0957-854X}
\affiliation{
  \institution{State Key Laboratory for Novel Software Technology, Nanjing University}
  \city{Nanjing}
  \state{Jiangsu}
  \country{China}
  \postcode{210093}
}

\author{Yuling Hu}
\email{yulinghu@smail.nju.edu.cn}
\orcid{0009-0001-0168-8842}
\affiliation{
  \institution{State Key Laboratory for Novel Software Technology, Nanjing University}
  \city{Nanjing}
  \state{Jiangsu}
  \country{China}
  \postcode{210093}
}

\author{Yuchen Chen}     \email{yuc.chen@outlook.com}
\orcid{0000-0002-3380-5564}
\affiliation{
  \institution{State Key Laboratory for Novel Software Technology, Nanjing University}
  \city{Nanjing}
  \state{Jiangsu}
  \country{China}
  \postcode{210093}
}
\author{Quanjun Zhang}    \email{quanjun.zhang@smail.nju.edu.cn}
\orcid{0000-0002-2495-3805}
\affiliation{
  \institution{State Key Laboratory for Novel Software Technology, Nanjing University}
  \city{Nanjing}
  \state{Jiangsu}
  \country{China}
  \postcode{210093}
}

\author{Xiuting Ge}    \email{dg20320002@smail.nju.edu.cn}
\orcid{0000-0003-3683-7374}
\affiliation{
  \institution{State Key Laboratory for Novel Software Technology, Nanjing University}
  \city{Nanjing}
  \state{Jiangsu}
  \country{China}
  \postcode{210093}
}

\author{Yang Liu}    
\email{yangliu@ntu.edu.sg}
\orcid{0000-0001-7300-9215}
\affiliation{
  \institution{School of Computer Science and Engineering, Nanyang Technological University}
  \state{50 Nanyang Avenue}
  \country{Singapore}
  \postcode{639798}
}

\author{Zhenyu Chen}
\email{zychen@nju.edu.cn}
\orcid{0000-0002-9592-7022}
\affiliation{
  \institution{State Key Laboratory for Novel Software Technology, Nanjing University}
  \city{Nanjing}
  \state{Jiangsu}
  \country{China}
  \postcode{210093}
}

\renewcommand{\shortauthors}{W. Sun, C. Fang, Y. Ge, Y. Hu, Y. Chen, Q. Zhang, X. Ge., Y. Liu., and Z. Chen.}


\begin{abstract}
(Source) code search is widely concerned by software engineering researchers because it can improve the productivity and quality of software development. Given a functionality requirement usually described in a natural language sentence, a code search system can retrieve code snippets that satisfy the requirement from a large-scale code corpus, e.g., GitHub. 
To realize effective and efficient code search, many techniques have been proposed successively. These techniques improve code search performance mainly by optimizing three core components, including query understanding component, code understanding component, and query-code matching component. 
In this paper, we provide a 3-dimensional perspective survey for code search. Specifically, we categorize existing code search studies into query-end optimization techniques, code-end optimization techniques, and match-end optimization techniques according to the specific components they optimize. 
Considering that each end can be optimized independently and contributes to the code search performance, we treat each end as a dimension. Therefore, this survey is 3-dimensional in nature, and it provides a comprehensive summary of each dimension in detail. 
To understand the research trends of the three dimensions in existing code search studies, we systematically review 68 relevant literatures. 
Different from existing code search surveys that only focus on the query end or code end or introduce various aspects shallowly (including codebase, evaluation metrics, modeling technique, etc.), our survey provides a more nuanced analysis and review of the evolution and development of the underlying techniques used in the three ends. 
Based on a systematic review and summary of existing work, we outline several open challenges and opportunities at the three ends that remain to be addressed in future work.
\end{abstract}



\maketitle

\section{Introduction}
\label{sec:introduction}
Software development is usually a repetitive task, where the same or similar implementations exist in established open-source projects (e.g., GitHub repositories~\cite{2008-GitHub}) or online forums (e.g., Stack Overflow~\cite{2008-Stack-Overflow}). Software developers on average spend about 19 percent of their development time in searching the relative code over a large-scale codebase, because they can reuse or modify previously written code snippets to improve the efficiency of project development~\cite{2009-Two-Studies-of-Opportunistic-Programming}. 
(Source) code search aims to retrieve relevant code in large-scale code corpora for a developer's given query requirements. Reusing the retrieved high-quality code can effectively improve the productivity and quality of software development, which makes code search receive widespread attention from software engineering researchers. 
Besides, code search provides more opportunities for code reuse, which is an earlier concept than code search~\cite{1984-Software-Reuse, 1987-software-reuse-through-information-retrieval, 2022-TranCS}. Code reuse aims to modify an existing code into a new code according to the requirements, while code search makes it possible to find the available one. In other words, code reuse can be realized through code search, further proving the considerable research value and significance of code search.

The most ideal code search can convert the query target input into the most relevant code snippet and return it to the user~\cite{2018-Effective-API-recommendation-without-historical-software-repositories, 2018-Deep-Code-Search}. In fact, many developers currently rely on Google, Baidu, and other search engines for code search, but the search results are arrestingly unsatisfactory. The search logic or methods used by common search engines are not suitable for matching natural language and code pairs. In most cases, the search results are not available for developers to refer to, which will take them more time to change or choose a better query~\cite{2019-DL-Met-CodeSearch, 2016-Query-Expansion-Based-on-Crowd-Knowledge-for-CS, 2018-Language-agnostic-Source-Code-Retrieval}. Given the importance of code search, there is great interest in implementing better methods for retrieving relevant code snippets from the code corpus, depending on the developer's intent expressed as a search query. 

Code search can be narrowly defined as a technique that takes as input a natural language query given by a user, then selects the closest or possible answer from the code corpus and returns it to the developer as an output. Of course, there exist a few individual code search techniques that take some non-natural language queries as input or return the output into other forms such as API interface~\cite{2011-Recommending-Proper-API-Code} or I/O examples~\cite{2016-Code-Search-with-Input-Output}. In this paper, we mainly study the mainstream code search scenario, where the query is natural language text and the search results are method/function code snippets. By organizing and conducting a statistical analysis of articles on code search that fit our main study, we can get a glimpse of the development history of code search techniques. Around the 1960s, the concept of code search emerged~\cite{1984-A-unified-approach-to-serial-search, 1989-An-Efficient-VQ-Code-Search-Algorithm, 1990-Service-Code-Search-High-Performance-Medical-Records}. Subsequently, with the development of open-source platforms, the number of papers related to code search has risen rapidly since 2009, and dozens of related papers appear in the field every year. Over the past nearly sixty years, many techniques have been successfully applied in code search, and there are also surveys and summaries of these techniques~\cite{2022-Survey-of-Code-Search-Tools, 2021-Survey-of-Automated-Query-Reformulations-in-CS, 2022-Code-Search-Survey}. 
These surveys either only focus on the query or code-end optimization or shallowly introduce various aspects of code search tools, including codebase, evaluation metrics, modeling techniques, etc.
For example, Rahman et al.~\cite{2021-Survey-of-Automated-Query-Reformulations-in-CS} explore the application of automatic composition query technology in code search, mainly reflected in the classification of algorithms, assessment of the quality of results and the future constraints and challenges of query reformulation. 
Another existing survey conducted by Liu et al.~\cite{2022-Survey-of-Code-Search-Tools} focuses on analyzing the existing code search tools, discusses and learns from the specific search content of the tools, and proposes relevant indicators for evaluating code search tools. In the latest survey completed by Luca et al.~\cite{2022-Code-Search-Survey} in 2022, the code search process, including query processing, code search models training, search results sorting and pruning are discussed and summarized.

However, a systematic review of various optimization techniques involved in different core components in code search techniques is still lacking. 
From the view of technical composition, both early and recent code search techniques are composed of three core components, including a query understanding component, a code understanding component, and a query-code matching component. The query understanding component and the code understanding component are responsible for mining and representing the features of the query and code snippet, respectively. The query-code matching component is responsible for ranking a set of candidate code snippets according to how closely their representations semantically match the query. 
The lack of a systematic review of the optimization techniques involved in these components hinders developers/researchers from identifying technology trends for each core component, as well as the challenges and opportunities faced in optimizing different components. 

In this survey, we propose a novel approach for summarizing code search techniques from a 3-dimensional perspective. 
Specifically, according to the specific components they optimize, we first classify existing code search techniques into three categories: query-end optimization techniques, code-end optimization techniques, and match-end optimization techniques. 
Each end exposes a perspective through which we can understand and dissect the essential optimizations/improvements made by existing code search techniques. 
Therefore, this survey investigates code search techniques from three perspectives, referred to as a 3-dimensional perspective. 
Then, we perform a systematic literature review of the optimization techniques proposed in the three ends. We carefully select 68 representative papers from the 1,427 candidates. Among these 44 papers propose solutions for query-end optimization, 50 papers for code-end optimization, and 52 for match-end optimization. Since some papers may propose optimization techniques for multiple ends, there will be some overlapping papers in the three ends. 
Finally, we tease out the development history of optimization techniques proposed for each end and summarize the development trends. 
Building on a thorough review of optimization techniques proposed in existing code search papers, we outline persistent challenges that necessitate further attention. Additionally, we present potential research opportunities in the code search field.

The main contributions of this survey include:
\begin{itemize}
   \item We propose a novel systematic review of 2,191 code search techniques published in journal papers, conference papers, and arxiv papers up to September 30, 2023, as a starting point for future code search research.
   \item We analyze 68 different code search techniques and conduct analysis from a 3-dimensional perspective to explore these techniques' innovations in the query-end optimization, code-end optimization, and match-end optimization. This survey will help subsequent researchers to enumerate their characteristics.
   \item We divide code search techniques into three categories according to the specific components they optimize, and analyze the evolution of techniques in each category over time as a basis for further comparison and benchmarking. 
   \item We highlight opportunities and challenges in code search research based on our findings to stimulate further research in this field. Some resources, datasets, and code can be found at \url{https://github.com/wssun/SourceCodeSearch}. 
\end{itemize}

\textbf{Structure of the paper:} The remainder of this survey is organized as follows. Section~\ref{sec:background} briefly introduces the background of code search. Section~\ref{sec:survey_methodology} presents the survey methodology that we follow. Section~\ref{sec:Answering_RQ1},~\ref{sec:Answering_RQ2}, and~\ref{sec:Answering_RQ3} summarize the key research questions we investigate and their answers in this study. Section~\ref{sec:outlook_and_challenges} discusses the challenges for the road ahead on code search techniques and presents the potential research opportunities for future work. Section~\ref{sec:threats_to_validity} shows the potential threats that may affect the validity of this review. Finally, Section~\ref{sec:conclusion} provides a conclusion of this survey.

\section{Background}
\label{sec:background}
In this section, we will introduce the background of code search, including code search definition and code search techniques.

\subsection{Code Search}
\label{subsec:code_search}
Since the existing works improve and implement the code search task using different techniques, there is no formal problem definition for the code search task. In this paper, we survey a wide range of code search research, including early information retrieval (IR)-based code search research and recent deep learning (DL)-based code search research. To the best of our knowledge, there is no strict definition of general code search. We investigate the widely-studied code search scenario, where the query given by the developer is a short natural language text, and the search result is a code snippet of a method/function~\cite{2009-Sourcerer, 2022-Survey-of-Code-Search-Tools, 2016-A-Search-Log-Mining-based-Query-Expansion, 2018-Deep-Code-Search, 2020-CodeBERT, 2021-Multimodal-Representation-for-Neural-Code-Search, 2022-TranCS}. Formally, let $q = \{w_1, w_2, \cdots, w_m\}$ be a query given by the developer, where $w_i$ is the $i$-th word in $q$; $S= \{s_1, s_2, \cdots, s_n\}$ be a large-scale code corpus, where $s$ is a code snippet; code search is defined as follows.

\begin{definition}[Code Search]
\label{def:code_search_task}
  Code search is the task of retrieving a code snippet $s \in S$ for $q$ that satisfies the following conditions:
\begin{itemize}
    \item $\forall s' \in S, s' \neq s$
    \item $\bm{q} = \Phi(q)$
    \item $\bm{s} = \Psi(s), \bm{s'} = \Psi(s')$
    \item $sim(\bm{q}, \bm{s'}) \leq sim(\bm{q}, \bm{s})$
\end{itemize}
where $\Phi(\cdot)$ and $\Psi(\cdot)$ are functions designed to represent features of query and code snippet, respectively. $sim(\cdot)$ is a function that measures the similarity between two feature representations of the query and code snippet.
\end{definition}

\begin{figure}[htbp]
  \centering
  \includegraphics[width=0.85\linewidth]{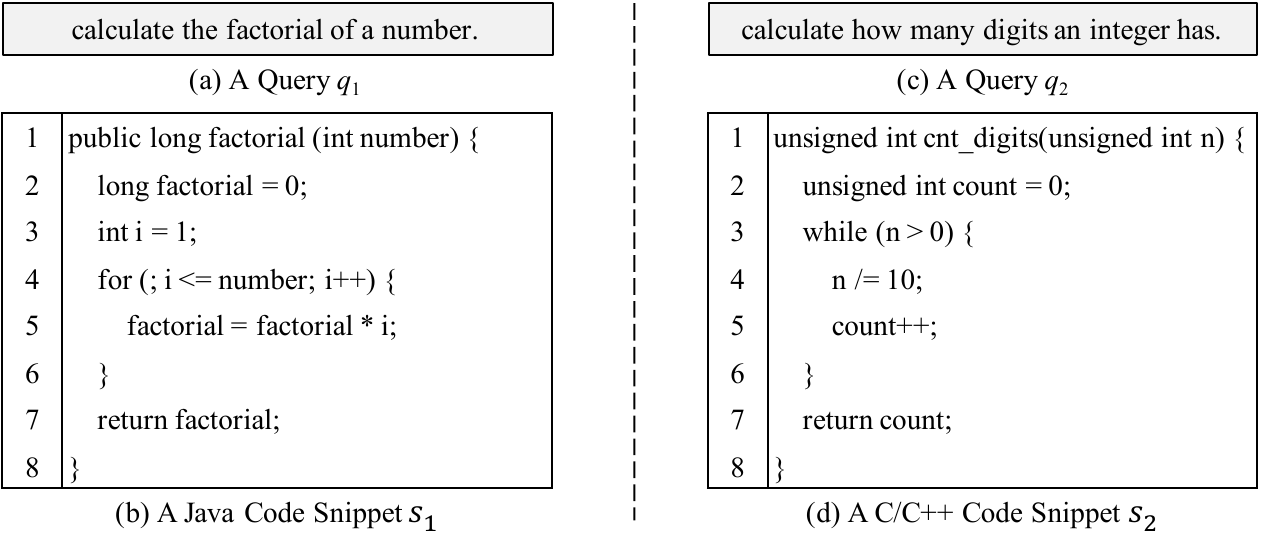}
  \caption{Examples of code search}
  \label{fig:example_of_code_search}
\end{figure}

We focus on investigating free-text code search where queries given by developers are free-form natural language text and search results are method-level code snippets. This is the most practical, common, and widely researched code search application scenario~\cite{2019-DL-Met-CodeSearch, 2022-Survey-of-Code-Search-Tools}. Figure~\ref{fig:example_of_code_search} shows two examples of code search. In both examples, queries are natural language descriptions and code snippets are methods written in programming languages (e.g., Java and C/C++). From Figure~\ref{fig:example_of_code_search}(a) and (c), we can observe that the former ($q_1$) wants to find a code snippet calculating the factorial of a number, while the latter ($q_2$) aims to retrieve a code snippet implementing the functionality of counting the number of digits in an integer value. Obviously, the two queries have different intents or requirements. The query understanding component is responsible for capturing the intents/semantics in the natural language query. Figure~\ref{fig:example_of_code_search}(b) and (d) show two code snippets. All code search systems require a code understanding component, which is responsible for capturing the semantics in the programming language code snippet. Intuitively, a good code search technique requires understanding the semantics of both the query and the code snippet. Only in this way, can it retrieve the code snippets that satisfy the query intents. 

\subsection{Code Search Techniques}
\label{subsec:code_search_techniques}

\begin{figure}[!t]
  \centering
  \includegraphics[width=\linewidth]{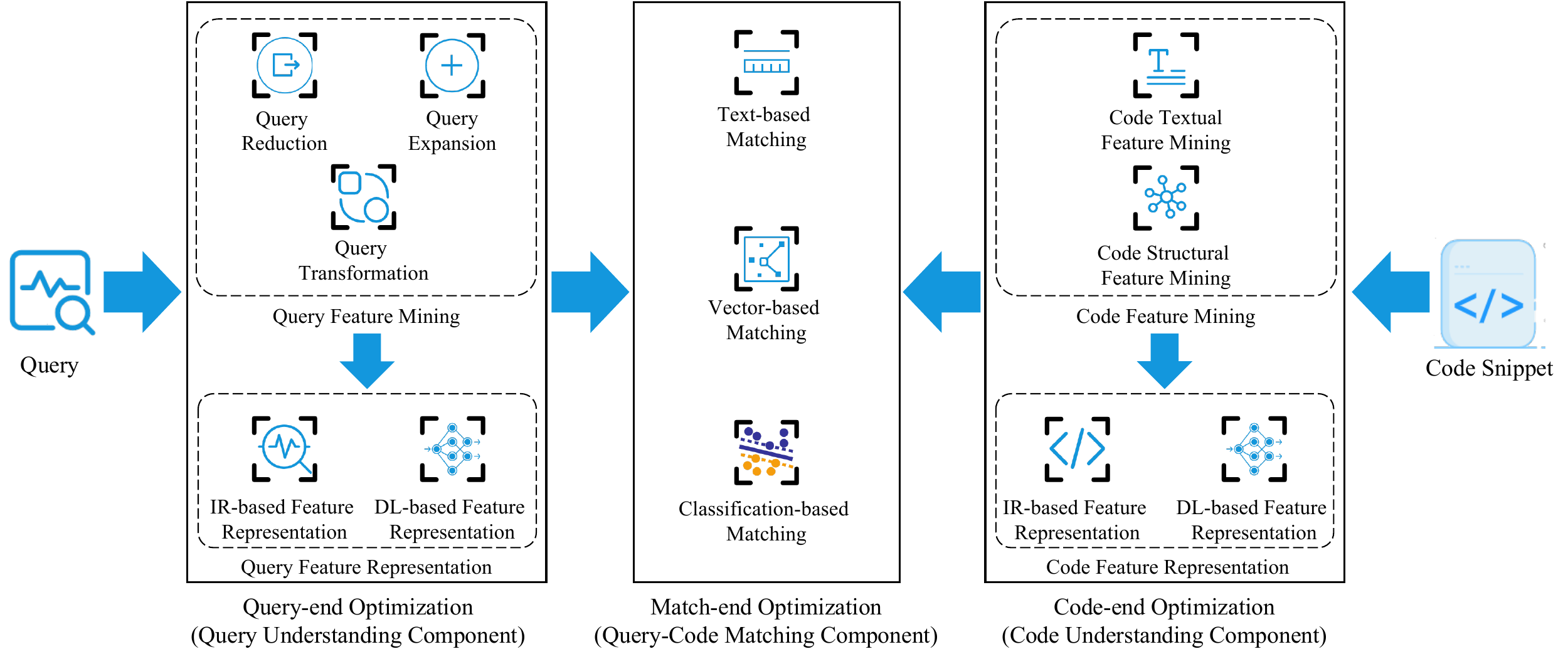}
  \caption{Overall framework of the code search technique}
  \label{fig:framework_of_code_search}
\end{figure}

Figure~\ref{fig:framework_of_code_search} shows the overall framework of the code search technique.
The goal of the code search technique is to retrieve relevant code from a large-scale code corpus according to the intent of the developers' query. 
A typical code search system/tool contains three components: a query understanding component, a code understanding component, and a query-code matching component.
The query understanding component is responsible for processing the natural language queries given by developers, such as mining and representing the key information (also called features/semantics) from the query. The code understanding component is responsible for processing the programming language code snippets in the code corpus, such as mining and representing features from the code snippets. The query-code matching component is responsible for ranking code snippets according to how well they semantically match the query. So, to implement effective and efficient code search, existing works have proposed various techniques to optimize these three components. In this survey, according to the specific components they optimize, we divide the existing code search studies into the following three categories: \textit{query-end optimization} techniques, \textit{code-end optimization} techniques, and \textit{match-end optimization} techniques.

\textbf{Query-end Optimization.} Given a query, query-end optimization aims to produce a query representation that not only preserves the core semantics of the query but also facilitates the computing of matching components, thereby improving the effectiveness and efficiency of code search. 
The query understanding component produces such a representation through two sequential steps: query feature mining and query feature representation. The query feature mining step treats the raw query given by the developer as input and extracts the important features from the raw query. Considering that the raw query may be of low quality and contains few useful features, existing works propose a variety of techniques to automatically optimize the quality of the features extracted from the raw queries~\cite{2015-Query-expansion-via-WordNet, 2016-Query-Expansion-Based-on-Crowd-Knowledge-for-CS, 2018-Interactive-Query-Reformulation, 2021-Survey-of-Automated-Query-Reformulations-in-CS, 2022-Enriching-Query-Semantics-for-CS-with-Reinforcement-Learning}. In this survey, we investigate four classic techniques for optimizing query quality, including query reduction, query replacement, query expansion, and query transformation (details are described in Section~\ref{subsec:query_feature_mining}). The query feature representation step takes in the features extracted by the query feature mining step and produces a semantic-preserving feature representation. Such a representation will be used to rank a large number of code snippets in the query-code matching component. 
From another perspective, query representation techniques are equally important. They also determine the final performance of the code search techniques, because accurate representation of semantic features in queries can assist code search models in retrieving code snippets correctly. 
Existing works propose many techniques to optimize the process of feature representation~\cite{2018-Retrieval-on-Source-Code-A-Neural-CS}. 
We divide the query representation methods utilized in existing technologies into two categories: IR-based feature representation and DL-based feature representation. 
IR-based representation methods always regard queries as vectors or plain texts, which can be utilized easily to optimize the query-end by the researchers.
DL-based feature representation methods apply deep neural networks to encode the query. They are frequently adopted by existing code search techniques because they yield better semantic representations of queries.
Details of query feature representation are described in Section~\ref{subsec:query_feature_representation}.

\textbf{Code-end Optimization. }
Given a code snippet, code-end optimization aims to produce a code representation that not only preserves the core semantics of the code but is also convenient for subsequent query-code matching computing, thereby improving the effectiveness and efficiency of code search.
Like the query understanding component, the code understanding component produces such code representation through two sequential steps: code feature mining and code feature representation. The code feature mining step treats the raw code snippet in the code corpus as input and extracts its important features.
Considering that the raw code snippet may be complex and contains noise features, existing works propose a variety of techniques to automatically optimize the quality of the features extracted from the raw code snippets~\cite{2018-Deep-Code-Search, 2018-Learning-to-represent-programs-with-graphs, 2019-DL-Met-CodeSearch, 2019-Multi-modal-Attention-for-Code-Rerieval, 2020-TranS3, 2021-DGMS, 2023-deGraphCS, 2022-Incorporating-Code-Structure-and-Quality-in-Deep-Code-Search}. It is a common practice to characterize code features from two aspects: textual features and structural features.
In this survey, we investigate three textual features and four structural features commonly used in code search works (details are described in Section~\ref{subsec:code_feature_mining}). The code feature will be further passed into the code representation step to produce semantic-preserving feature representations. Such a representation will be used to rank a large number of code snippets in the query-code matching component. To make the resulting representation as semantically preserving as possible, existing works propose many techniques to optimize the process of code feature representation~\cite{2018-Retrieval-on-Source-Code-A-Neural-CS}. 
Like query feature representation, we also divide code feature representation techniques into two categories: IR-based feature representation and DL-based feature representation. 
IR-based feature representation treats the code primarily as text, supplemented by additional information to represent it more comprehensively.
DL-based feature representation techniques apply deep neural networks to embed the code features. Details of code feature representation are described in Section~\ref{subsec:code_feature_representation}.

\textbf{Match-end Optimization.} Given a query representation and a set of code representations, match-end optimization aims to rank code representations based on their relevance to the query representation. Different techniques adopt different methods to calculate the relevance scores. In this survey, we investigate three widely used matching methods, including text-based matching, vector distance-based matching, and classification-based matching. Text-based matching methods measure the relevance scores by calculating the distance between the keyword-based or IR-based query feature representation and the keyword-based or IR-based code feature representation. Vector-based matching methods measure the relevance scores by calculating the distance between the query feature vectors (including embeddings) and the code feature vectors. Vectors are produced by traditional IR techniques or advanced DL techniques. Classification-based matching methods measure the relevance scores by using neural network classifiers to predict the probability of semantic relevance of the query embeddings and the code embeddings. 

\section{Survey Methodology}
\label{sec:survey_methodology}
We follow the guidelines for the systematic literature review (SLR) in software engineering~\cite{2007-Guidelines-for-SLR-in-SE, 2013-Systematic-Review-in-SE, 2015-Guidelines-for-SMS-in-SE} to conduct this survey. We start this survey by asking three research questions and then comprehensively analyze the various optimization techniques used in code search.

\subsection{Research Questions}
\label{subsec:research_questions}
The research questions are one of the most important contents of the literature review, which guide us to clarify the research direction and thus conduct a purposeful investigation. In this survey, we want to explore, classify, and summarize the various optimization techniques used in code search to date in query-end optimization, code-end optimization, and match-end optimization. These three dimensions are the core components of the code search techniques and the main perspectives for researching code search techniques in this survey. Therefore, we investigate the following three research questions (RQs): 

\begin{itemize}
    \item \textbf{RQ1.} What are the query-end optimization techniques in code search studies? The purpose of this RQ is to investigate which techniques are applied to query processing (including query feature mining and query feature representation) and the development trend of query-end optimization techniques.
    
    \item \textbf{RQ2.} What are the code-end optimization techniques in code search studies? This RQ aims to investigate which code features are used in code-end optimization techniques, how they represent involved code features, and the development trend of code-end optimization techniques.
    
    \item \textbf{RQ3.} What are the match-end optimization techniques in code search studies? The goal of this RQ is to investigate how match-end optimization techniques match the representations of the query and code features to rank the expected code snippet in a higher rank and the development trend of match-end optimization techniques.
\end{itemize}

In the process of seeking the answers to these questions, we also find the deficiencies and limitations of the existing code search techniques. As a reference, we put forward some suggestions and directions for future research.

\begin{figure}[!t]
  \centering
  \includegraphics[width=\linewidth]{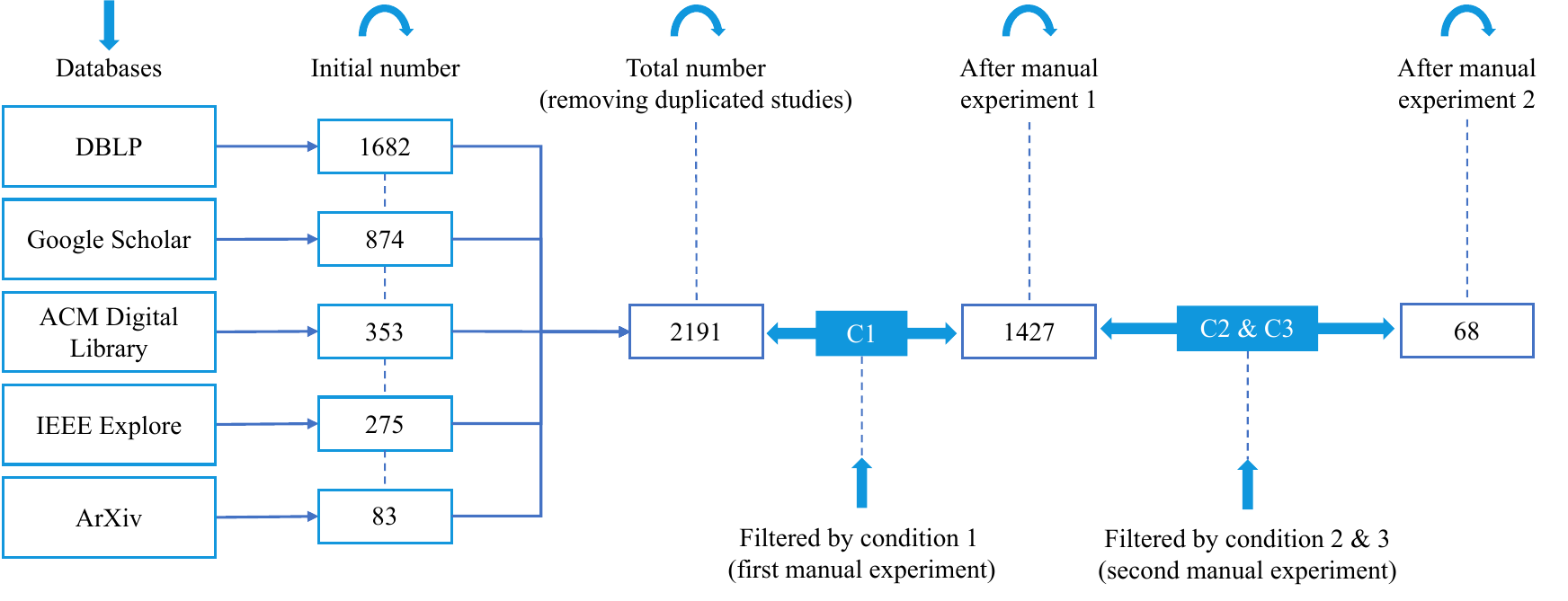}
  \caption{Selection of research studies}
  \label{fig:selection_of_research_studies}
\end{figure}

\subsection{Search Strategy}
\label{subsec:search_strategy}
Collecting published papers requires selecting appropriate publication databases and searching keywords. In this survey, we selected six widely available electronic databases, including DBLP publication database~\footnote{\url{https://dblp.uni-trier.de}}, Google scholar database~\footnote{\url{https://scholar.google.com/}}, IEEE Explore database~\footnote{\url{https://ieeexplore.ieee.org/}}, ACM Digital Library~\footnote{\url{https://dl.acm.org/}}, Web of Science database~\footnote{\url{https://www.webofscience.com/}}, and ArXiv database~\footnote{\url{https://arxiv.org/}}. It is worth noting that, considering that some researchers are willing to disclose the latest research techniques on ArXiv in advance, we also collected the code search papers that have been made public on ArXiv but not accepted by any journal or conference. 
In terms of selecting searching keywords, we referred to the \textbf{PIO} (Population + Intervention + Outcome) criteria~\cite{2013-A-systematic-reviewo-of-systematic-review-process-research} and used the population terms and intervention terms from code search field to construct keywords, as in previous studies~\cite{2021-Survey-of-Automated-Query-Reformulations-in-CS}. \textbf{Population terms} covered all aspects of the research topic. In this survey, we used ``Code Search'', ``Code Retrieval'', ``Code Recommendation'', and ``Code Reuse'' as the population terms. \textbf{Intervention terms} focus on a specific aspect of the research topic. We used ``Query Expansion'', ``Query Reduction'', and ``Query Transformation'' as the intervention terms for RQ1. 
Relevant papers up to September 30, 2023, were included in our search results. 
As Figure~\ref{fig:selection_of_research_studies} shows, we retrieved 2,191 relevant papers from six databases in total as \textbf{Outcome} after removing duplicated studies.

\subsection{Study Selection}
\label{subsec:study_selection}
Once those candidate studies were collected, we made a gradual selection according to the filter conditions (\textbf{C}) we carefully set up. 

\begin{itemize}
    \item \textbf{C1.} eliminating book, thesis, and short papers (our definition of short essay papers: less than five pages); only needing journals and conferences.
    \item \textbf{C2.} only selecting technical papers, excluding technical reports, empirical studies, and surveys.
    \item \textbf{C3.} choosing the papers about the field of code search, and also needing to meet the requirements that have a technical innovation in one of the ends.
\end{itemize}

According to the three conditions, manual experiments were carried out to select the final range from the candidate papers. The first manual experiment obtained the page number and published source of the whole paper. We obtained a total of 1,427 papers (including 708 conference papers, 645 journal papers, and 74 arxiv papers) according to the requirements of Condition \textbf{C1}. In the second manual experiment, we judged whether the remaining papers meet Conditions \textbf{C2} and \textbf{C3} based on the title, abstract, and approach part.

\begin{figure*}[htbp]
  \centering
  \includegraphics[width=\linewidth]{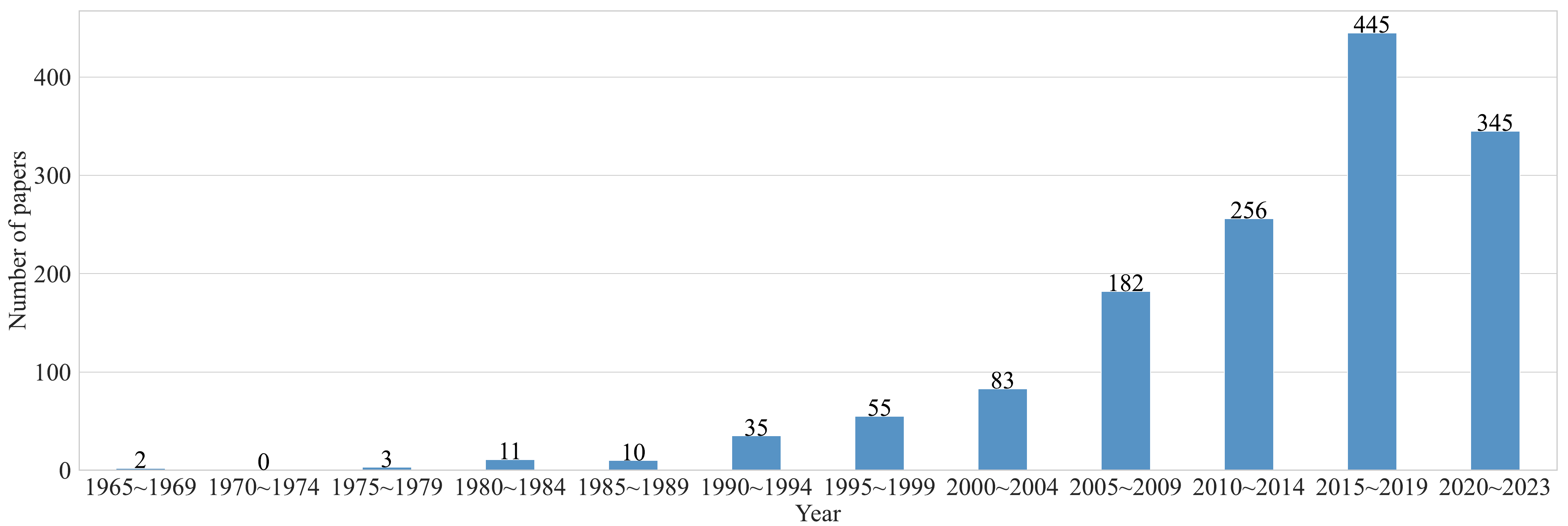}
  \caption{1,427 code search papers from 1965 to 2023. Please replace the data in this figure with our data.}
  \label{fig:paper_trend_by_year}
\end{figure*}

We grouped the collected 1,427 papers by year of publication, and the statistical results are shown in Figure~\ref{fig:paper_trend_by_year}. 
It is observed that the number of relevant papers has increased significantly since 2005, indicating that the problem of code search has received significant attention.

\begin{figure*}[!t]
    \centering
    \subfigure[Group by Year]
    {
        \includegraphics[width=0.45\linewidth]{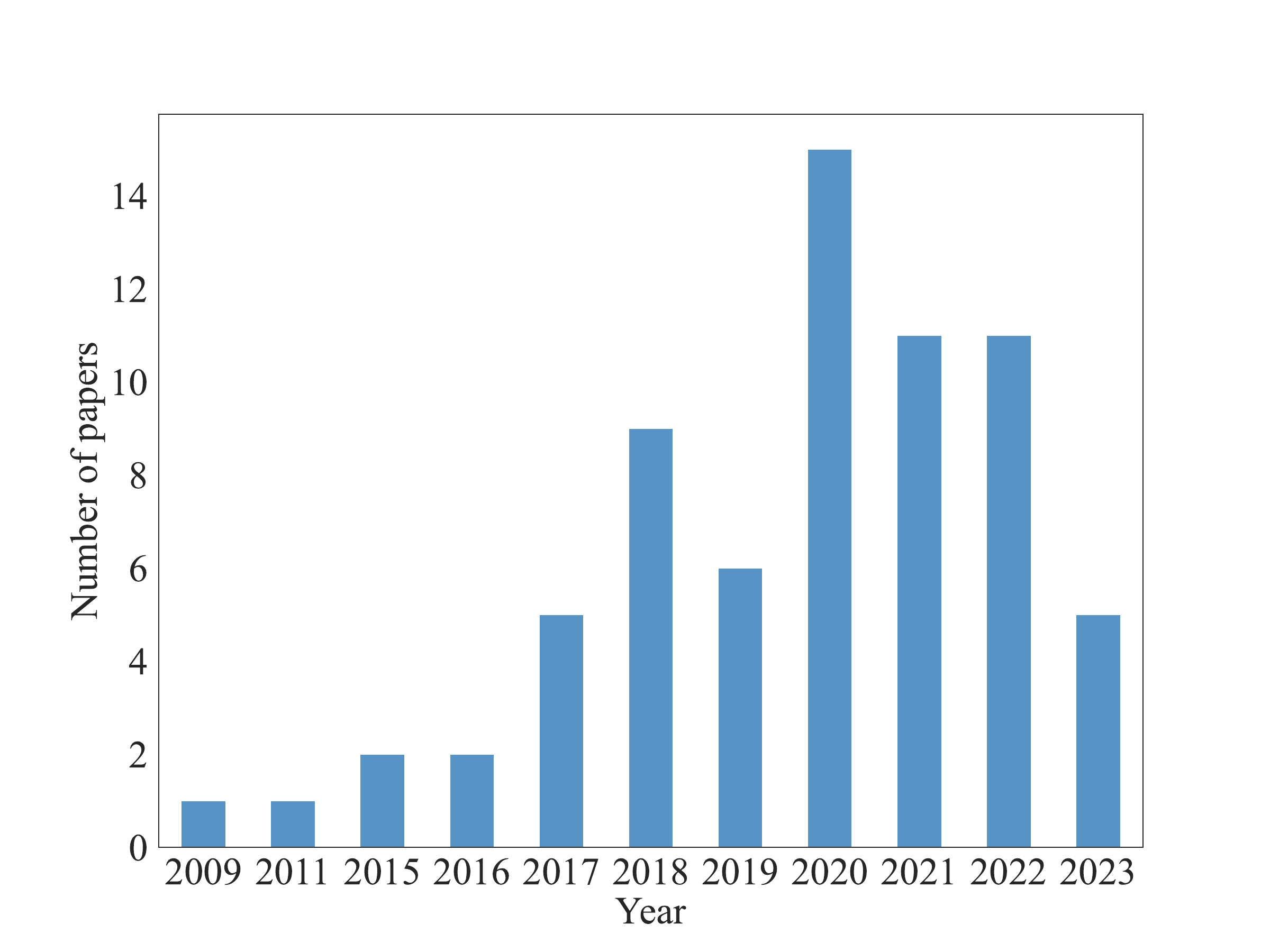}
        \label{fig:paper_we_investigate_by_year}
    }
    \hspace{0.1\linewidth}
    \subfigure[Group by Venue]
    {
        \includegraphics[width=0.35\linewidth]{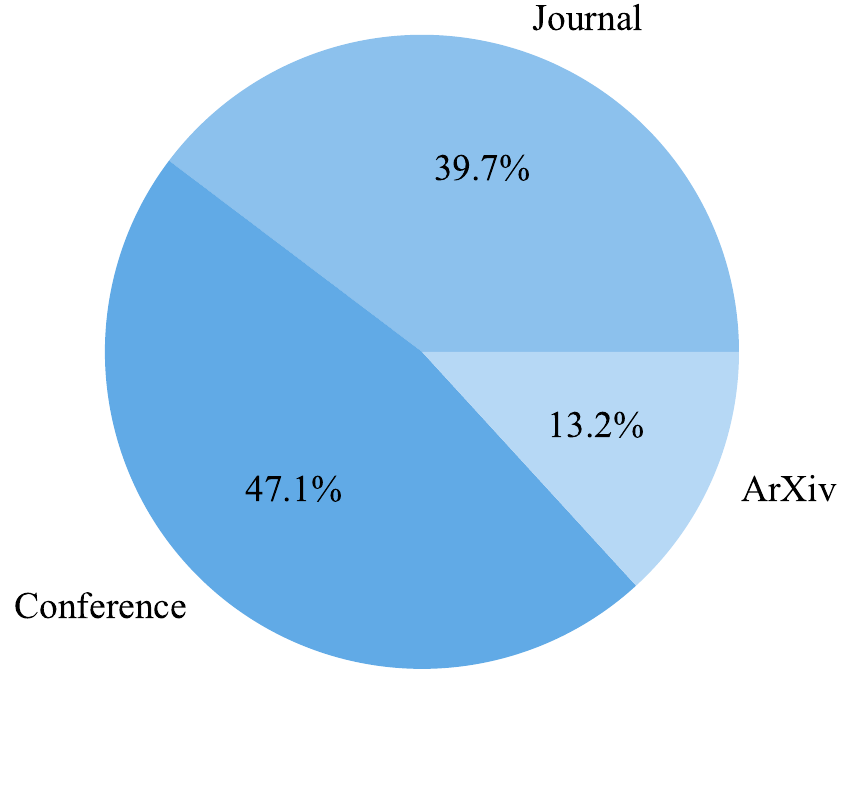}
        \label{fig:paper_we_investigate_by_venue}
    }
    \caption{68 papers on code search discussed in this article}
    \label{fig:paper_we_investigate}
\end{figure*}

After screening all the conditions, we finally identified a total of code search studies related to our research. For this study, we carefully selected representative papers from the candidates and conducted a systematic analysis, and finally, 68 studies are selected. According to the three questions we raised above, these papers are divided into three relevant categories, among which 44 papers are related to RQ1; 50 papers are related to RQ2; and 52 papers are related to RQ3. Since some papers may propose optimization techniques for multiple ends, there will be some overlapping papers in the three RQs. It should be noted that we will focus on introducing the optimization techniques proposed by each paper on the corresponding end in different RQs. Figure~\ref{fig:paper_we_investigate_by_year} presents the number of papers we discuss per year of publication, illustrating the increasing relevance of the topic. Figure~\ref{fig:paper_we_investigate_by_venue} shows that 47.1\% of them were published in conference proceedings, accounting for the largest proportion, followed by Journal (39.7\%) and ArXiv (13.2\%).

\section{Answering RQ1: What are the query-end optimization methods in code search studies?}
\label{sec:Answering_RQ1}
In the realm of code search, the quality of a query is paramount as it directly influences the quality of the search results~\cite{2016-RACK, 2019-Automatic-Query-Reformulation-for-CS}. 
A low-quality query, characterized by the use of uncommon abbreviations or redundant and noisy phrases, may result in the retrieval of suboptimal code snippets. As a countermeasure, researchers have progressively proposed a variety of optimization techniques aimed at optimizing and understanding user queries. These endeavors collectively serve to enhance the performance of code search. 
In a nutshell, query-end optimization endeavors to elevate the quality of user queries, consequently heightening the quality of code search.

\begin{figure*}[!t]
    \centering
    \begin{minipage}[l]{\linewidth}
      \resizebox{0.98\textwidth}{!}{
        \begin{forest}
            for tree={
                grow=east,
                reversed=true,
                anchor=base west,
                parent anchor=east,
                child anchor=west,
                base=left,
                font=\large,
                rectangle,
                draw=black,
                rounded corners,
                align=left,
                minimum width=4em,
                edge+={darkgray, line width=1pt},
                s sep=3pt,
                inner xsep=2pt,
                inner ysep=3pt,
                line width=1pt,
            },
            where level=1{text width=10em, font=\normalsize}{},
            where level=2{text width=4.5em, font=\normalsize}{},
            where level=3{text width=4.5em, font=\normalsize}{},
            where level=4{text width=4.5em, font=\normalsize}{},
            where level=5{text width=4.5em, font=\normalsize}{},
            where level=6{text width=4.5em, font=\normalsize}{},
            where level=7{text width=4.5em, font=\normalsize}{},
            where level=8{text width=4.5em, font=\normalsize}{},
            where level=9{text width=4.5em, font=\normalsize}{},
            [Query\\Feature\\Mining
                [\hspace{9.5pt}Query Reduction \\ \hspace{22pt}(\S\ref{subsubsec:query_expansion})
                    [
                      [\hspace{6.5pt}\cite{2016-APIBook}
                        [
                          [
                            [\hspace{7.5pt}\cite{2019-QESC}  
                              [
                                [
                                  [
                                    [\hspace{6.5pt}\cite{2023-XCoS}]]]]]]]]]
                ]
                [\hspace{9.5pt}Query Expansion \\ \hspace{22pt}(\S\ref{subsubsec:query_expansion})
                    [\cite{2015-CodeHow, 2015-Query-expansion-via-WordNet}
                        [\hspace{7.5pt}\cite{2016-Query-Expansion-Based-on-Crowd-Knowledge-for-CS}
                            [\hspace{6.5pt}\cite{2017-IECS}
                                [\cite{2018-Interactive-Query-Reformulation, 2018-Expanding-Queries-for-Code-Search}\\
                                \cite{2018-Language-agnostic-Source-Code-Retrieval, 2018-Query-Expansion-Based-on-Code-Changes}
                                    [\cite{2019-NQE, 2019-QESC}
                                        [\hspace{7.5pt}\cite{2020-Unsupervised-Software-Repositories}
                                            [\hspace{10.5pt}\cite{2021-Search-for-Compatible-Code}
                                                [
                                                    [\hspace{7.5pt}\cite{2023-Self-Supervised-Query-Reformulation}
                                                    ]]]]]]]]]
                ]
                [\hspace{1.5pt} Query Transformation \\ \hspace{22pt}(\S\ref{subsubsec:query_transformation})
                    [
                        [
                            [
                                [\hspace{7.5pt}\cite{2018-Augmenting-and-Structuring-User-Query}
                                    [
                                        [
                                            [\hspace{7.5pt}\cite{2021-DGMS}
                                                [\hspace{6.5pt}\cite{2022-Enriching-Query-Semantics-for-CS-with-Reinforcement-Learning}
                                                    [\hspace{6.5pt}\cite{2023-Hyperbolic-Code-Retrieval}
                                                    ]]]]]]]]]
                ]
            ]
        \end{forest}
      }
    \end{minipage}
    \begin{minipage}[l]{\linewidth}
    \begin{tikzpicture}[x=30]
        \draw[line width=1pt] (0,0) -- (1.5,0);
        \draw[snake=snake, line width=1pt, segment amplitude=2pt, segment length=6pt] (1.5,0) -- (3,0);
        \draw[-stealth, line width=1pt,] (3,0) -- (13,0);
        
        \foreach \x in {3.9,5,6.2,7.4,8.6,9.7,10.8,12,13.1}
        \draw (\x cm,3pt) -- (\x cm,-3pt);
        
        \draw (0.9,0) node[below=3pt] {(year)} node[above=3pt] {};
        \draw (1.8,0) node[below=3pt] {} node[above=3pt] {};
        \draw (2.8,0) node[below=3pt] {} node[above=3pt] {};
        \draw (3.7,0) node[below=3pt] {2015} node[above=3pt] {};
        \draw (4.7,0) node[below=3pt] {2016} node[above=3pt] {};
        \draw (5.9,0) node[below=3pt] {2017} node[above=3pt] {};
        \draw (7.0,0) node[below=3pt] {2018} node[above=3pt] {};
        \draw (8.1,0) node[below=3pt] {2019} node[above=3pt] {};
        \draw (9.2,0) node[below=3pt] {2020} node[above=3pt] {};
        \draw (10.3,0) node[below=3pt] {2021} node[above=3pt] {};
        \draw (11.4,0) node[below=3pt] {2022} node[above=3pt] {};
        \draw (12.4,0) node[below=3pt] {2023} node[above=3pt] {};
        
    \end{tikzpicture}
    \end{minipage}
    
    \caption{Evolution of feature mining techniques in query end}
    \label{fig:evolution_query_feature_mining}
\end{figure*}
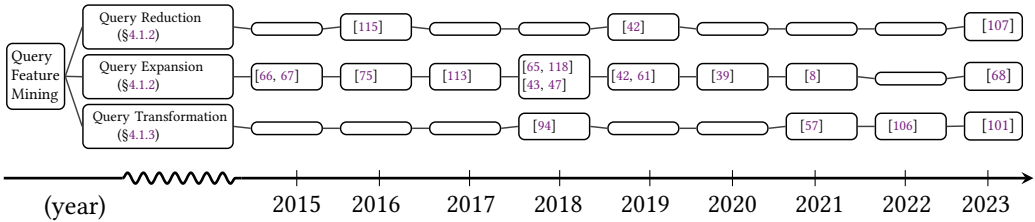

As shown in Figure~\ref{fig:framework_of_code_search}, the optimization techniques employed at the query end can be mainly categorized into two facets: feature mining from the user query and subsequent representation of these extracted query features. 
For query feature mining, existing efforts predominantly fall within three categories: query reduction, query expansion, and query transformation. 
Query reduction improves the quality of the user query by eliminating redundant content~\cite{2019-NQE}. 
Query expansion is to enrich the user query by incorporating available information, such as software-specific expansion words sourced from Stack Overflow~\cite{2015-Query-expansion-via-WordNet, 2016-Query-Expansion-Based-on-Crowd-Knowledge-for-CS}. 
Query transformation first transforms the user query into an alternative form, such as API~\cite{2016-SWIM}, and then retrieves related code snippets using the transformed form. 
Figure~\ref{fig:evolution_query_feature_mining} lists some representative works of the three types of query feature mining techniques at various temporal junctures. More details of these techniques are discussed in Section~\ref{subsec:query_feature_mining}. 
As for query feature representation, existing endeavors are generally classified into two classes: IR-based methods and DL-based methods. The main difference between the two lies in the adoption of different techniques to embed the natural language4 query and generate the corresponding vector representations. 
Figure~\ref{fig:evolution_query_feature_representation} presents some notable works of the two categories of query feature representation techniques across different points in time. More details of query feature representation are discussed in Section~\ref{subsec:query_feature_representation}.
In the following subsections, we will intricately introduce the three types of query feature mining techniques and the two categories of query feature representation techniques aforementioned in detail. Additionally, we will succinctly encapsulate their trajectories of development and emerging trends.

\subsection{Query Feature Mining} 
\label{subsec:query_feature_mining}

\subsubsection{Query Reduction}
\label{subsubsec:query_reduction}
As mentioned earlier, query reduction strives to eliminate superfluous components from the query, which contributes to the code search technique to grasp the core intent of the query accurately, thereby enhancing performance. 
Existing studies~\cite{2013-Query-Reformulation-for-Text-Retrieval, 2016-QUICKAR} find that the keywords that occur in more than 25\% of the documents in a corpus are less discriminating. Thereby, the key of query reduction is to remove the redundant, noisy, ambiguous, or less discriminating keywords~\cite{2021-Survey-of-Automated-Query-Reformulations-in-CS}, making the results of code search more reasonable and suitable for users to reuse. The first line of Figure~\ref{fig:evolution_query_feature_mining} shows the code search papers that employ query reduction to reformulate user queries. It can be seen intuitively that query reduction is less common for code search. In the following, we will detail how these two code search works perform query reduction.

Yu et al.~\cite{2016-APIBook} propose an information retrieval-based technique called APIBook to help users search the code of API methods. Given a user query also called ``API description'' in their paper, APIBook extracts semantic information and type information from the query. The extracted information will be used for matching later. The semantic information of an API description refers to the meaning of words in the API description. They use nouns, verbs, and adjectives as the semantic information of the API description. The type information of an API description refers to the information that concerns types, such as ``String'' and ``StringBuilder''. In simple terms, they perform query reduction by removing content other than semantic and type information, such as stop words, prepositions, and adverbs.

Huang et al.~\cite {2019-QESC} propose a deep learning-based method called QESC for effective query expansion. Although only query expansion is emphasized in the title of the paper, QESC also performs query reduction. Therefore, it can be said that QESC employs a combination of query expansion and query reduction to optimize query feature mining. In this section, we focus on how QESC performs query reduction, and the content of query expansion is introduced in Section~\ref{subsubsec:query_expansion}. Specifically, they decompose the code search into two steps: first-pass retrieval and second-pass retrieval. In the first-pass retrieval, on receiving a query, the search engine produces the initial query results.
They train an inference model based on the Deep Belief Network~\cite{2006-Deep-Belief-Nets}, which can infer the fine-grained changed code terms that will most likely occur in the initial results. 
They represent a code term by a triplet of ($\langle label\rangle, \langle role\rangle, \langle operation\rangle$). $\langle label\rangle$ represents the textual information of Abstract Syntax Tree (AST) nodes. $\langle role\rangle$ has two options that decide if a term is changed or dependent. $\langle operation\rangle$ has three operations that decide if a term is unchanged, new, or deleted. 
In the second-pass retrieval, they reformulate the query with the selected changed terms. If the $\langle operation\rangle$ of the term is ``deleted'', they see it as an irrelevant term and remove its $\langle label\rangle$ from the query. 
In short, QESC performs query reduction by deleting irrelevant terms identified by the inference model.

Wang et al.~\cite{2023-XCoS} find the existing code search tools usually return a ranked list of candidate code snippets without any explanations, making the developers often find it hard to choose the desired results and build confidence on them. 
To address this issue, they propose XCoS, an explainable code search approach based on query scoping and knowledge graph. 
Query scoping essentially performs the process of query reduction, aiming to extract different parts from a query, including functionalities, functional constraints, and nonfunctional constraints. 
Specifically, given a query, XCoS first removes the starting words for a question, such as ``how to'' and ``how can I''. Then XCoS uses an NLP tool spaCy to analyze the part of speech and dependence tree of the remaining query. After that XCoS extracts different parts from the query using linguistic rules. The linguistic rules for the code search query are borrowed from the NLP field where linguistic rules are commonly used to extract functionality and constraints~\cite{2008-Using-Linguistic-Knowledge-Classify-Non-functional-Requirements, 2015--Extracting-Development-Tasks-to-Navigate-Documentation, 2016-ICON}.
To build linguistic rules, they randomly sample 50 question titles as a validation dataset to iteratively refine and validate the rules by observing the extraction results. In the rules, VERB, DOBJ, PREP, POBJ, and MOD denote verb, direct object, preposition, preposition object, and modifier, respectively. Based on these rules, they summarize that \textit{Functionality: VERB or VERB DOBJ; Functional Constraint: PREP POBJ, to VERB DOBJ, or using DOBJ; Nonfunctional Constraint: adverbs, adverb clauses, relative clauses, or phrases such as in MOD way that are used to modify or qualify the functionalities or functional constraints}. XCoS composes the functionality part and functional constraint part together to retrieve code snippets.
The knowledge graph is used to guide the generation of explanations for code snippets and its construction is an offline task. 
In practice, given a query, XCoS first identifies different parts (i.e., functionalities, functional constraints, nonfunctional constraints) from it and uses the expressions of functionalities and functional constraints to search the codebase. It then links both the query and the candidate code snippets to the concepts in the knowledge graph and generates explanations based on the association paths between these two parts of concepts together with relevant descriptions.

\subsubsection{Query Expansion}
\label{subsubsec:query_expansion}
Query expansion is to expand the user's original query through the inclusion of other available information, such as synonymous words. This approach serves as a strategy for reformulating queries, particularly in cases where the original query yields a poor retrieval result. Originally, in the field of information retrieval, query expansion was used for cross-language information retrieval~\cite{2007-Cross-Language-Query-Expansion}, the words/terms suggested by query expansion techniques can rich retrieval code results. Given the effective promotion of search performance, the researchers in the software engineering community introduce them to code search tasks. The second line Figure~\ref{fig:evolution_query_feature_mining} presents the code search papers that leverage query expansion to enrich user queries. It is apparent that, compared with query reduction, query expansion has garnered extensive attention and application within code search. Subsequently, we will discuss how these code search endeavors conduct query expansion.

Lv et al.~\cite{2015-CodeHow} find that one major limitation of existing code search tools is the lack of query understanding. These tools often adopt conventional text similarity matching techniques to retrieve relevant code snippets. They do not consider query understanding, which could lead to inaccurate return results. Therefore, they propose CodeHow, a code search approach that considers both API understanding and text similarity matching. CodeHow understands a query by identifying the APIs that the query may refer to. CodeHow expands the user query with the identified APIs and applies the Extended Boolean model to retrieve the code snippets that match the expanded query. Specifically, CodeHow decomposes code search into two phases, i.e., the API understanding phase and the code retrieval phase. 
In the API understanding phase, CodeHow first collects the description of each API in the API library from its online documentation. It then calculates two similarity values, one between the API description and the query, and one between the API name and the query. Finally, it returns the potentially relevant APIs that match the query according to two similarity values. 
In the code retrieval phase, CodeHow constructs a Boolean query expression for retrieving code snippets that match the query in terms of text similarity. It retrieves code snippets that contain the potentially relevant API as well as other query terms in the method body and method name.
CodeHow also constructs Boolean query expressions for each API recommended in the API understanding phase, which are intended to search for code snippets that contain the potentially relevant API. The query expressions above are combined to obtain an expanded query expression for retrieving code snippets. 
Finally, the expanded query expression is passed to an Extended Boolean Model as the input, which will return relevant code snippets according to their similarity to the expanded query expression.

To overcome the term mismatch problem inherent in text retrieval-based techniques, Nie et al.~\cite{2016-Query-Expansion-Based-on-Crowd-Knowledge-for-CS} propose Query Expansion based on Crowd Knowledge (QECK) to improve the performance of code search. Specifically, given a query, QECK ranks all Question \& Answer (Q\&A) pairs collected from Stack Overflow~\footnote{\url{https://stackoverflow.com/}} using the information retrieval model Lucene~\footnote{\url{ http://lucene.apache.org}}. The top-$m$ Q\&A pairs are identified as the Pseudo Relevance Feedback (PRF) documents, which will be treated as relevant to the query. Then, QECK identifies useful expansion words from PRF Q\&A pairs. Each word in PRF Q\&A pairs is assigned an expansion weight. According to the weights of words, top-$n$ words are selected as useful expansion words and added to the original query to generate the expanded queries. Finally, QECK ranks all code snippets in the corpus for the expanded query, and the top $k$ code snippets are recommended to developers as the search results.

Yang et al.~\cite{2017-IECS} discover a significant issue where code search results are often modified manually. This phenomenon is caused by the inability to predict intent accurately with code search tools. To address this problem, they propose an intent-enforced code search approach called IECS. IECS can predict potential intents for a query before performing code retrieval. It utilizes intent to enhance the search to meet user needs. 
Specifically, IECS extracts intents from the given query by the intent extraction algorithm. This algorithm first uses the AST to identify modifications from the past method records. It then records the mapper between the identifiers and the concrete instances according to the modifications. After that, IECS can extract the intents from the mapper. Finally, it expands the query with intents and applies the Extended Boolean Model to retrieve the relevant code without any subsequent modification.

Zhang et al.~\cite{2018-Expanding-Queries-for-Code-Search} find that the proportion of identifiers (e.g., class and method names) plays a key role in retrieving relevant code examples from code search engines. Therefore, they propose to perform query expansion by recommending semantically related identifiers (particularly Application Program Interface (API) class-names) to expand natural-language queries. 
Specifically, they first use the continuous bag-of-words model (CBOW)~\cite{2013-Efficient-Estimation-of-Word-Representations-in-Vector-Space} to extract vector representations that are used to map natural-language queries with identifiers. 
Then based on similarities between vector representations of the query and API class-names, they find relevant API class-names (identifiers) from the corpus to expand the given query.
Finally, they retrieve code snippets from the corpus by executing the expanded query.

Lu et al.~\cite{2018-Interactive-Query-Reformulation} find that formulating (e.g., exchanging, adding, and deleting) the related words identified based only on the positional proximity is not enough to optimize the query. The presence of word relations prevalent within the source code, such as compound words and synonyms, are useful for query reformulation. Therefore, they propose a novel method called INQRES to interactively leverage these word relations to expand the query. 
Specifically, given a query, INQRES first extracts meaningful keywords from it. Then INQRES expands the keywords based on identifiers from the source code, meaningful words from the comments, and synonyms from the WordNet thesaurus~\cite{1998-Wordnet}. 
To find related expansion words from the three sources, INQRES excavates five word relations in the source code, including the Inheritance Relation (InR), Implementation Relation (ImR), Synonym Relation (SynR), Same-word Relation (SamR) and Compound Relation (ComR). All extracted word relations are saved in a word set called word-relation library (WRLib). Then, WRLib is used to extend the related words in the search query. If some words in the query also occur in WRLib, the related words are recommended, which are annotated by the relation level (SynR-1, SynR-SamR-2, etc.), which consists of the type of word relations and the grade of the relation between the recommended words and the query words. All extended related words are sorted based on the grade of relation level and their frequency used in the source code. 
After ranking all the related words, they further demonstrate them using the ``AND'' or ``OR'' relation for developers to understand and select.
Inspired by code search engines where the developers often use the ``advanced search'' to optimize the query results, that is, query words are combined with the ``AND'' and ``OR'' relations, INQRES builds ``AND'' and ``OR'' relations in an interactive way for the developer to select suitable words for query expansion. SynR and SamR indicate similar relations between words and thus the words in these two sets are defined as the ``OR'' relation. The meanings of ComR, InR, and ImR are complementary, and thus the words in these two sets are defined as the ``AND'' relation. 
In INQRES, all related words are shown in an interface. Users are empowered to evaluate the relevance of these words to the original query. Those deemed relevant are recognized as valid and subsequently incorporated into the original query. INQRES can iteratively use the expanded query to identify other effective related words in a similar way, until the search results are satisfied.

Code tokenization is a key preprocessing step in code search techniques, which aims to convert query or code snippet into lexicons. 
Karnalim~\cite{2018-Language-agnostic-Source-Code-Retrieval} find that most code search techniques rely on programming-language-dependent features to extract source code lexicons. However, these techniques would require manual updates to accommodate new programming languages, a process that can consume a significant amount of time. 
To handle this issue, Karnalim proposes a language-agnostic code retrieval approach. It does not rely on programming-language-dependent features.
Instead, it relies on the Keyword \& Identifier lexical pattern which are typically similar across various programming languages. The recognized lexicons are classified by lexicon categorization based on Keyword \& Identifier lexical pattern. This lexical pattern is selected as the main concern, because its rules are similar in most programming languages. 
This pattern is also adapted to query expansion. Specifically, the keyphrases found on the most descriptive paragraph are regarded as the query expansion candidates. They are selected from top-K retrieved documents and limited by lexical pattern. A lexicon is only viewed as a candidate if its category is similar to the category of the query term, either a keyword-like or identifier-like lexicon. After that, the candidates are sorted by the importance score which is weighted by term frequency and one-to-many association. Finally, the query expansion candidates are used to improve the code search performance, which expands the query based on terms found in top-K retrieved source codes. 

NCS~\cite{2018-Retrieval-on-Source-Code-A-Neural-CS} is a useful code search tool that can correctly search repositories of existing source code for code snippets. However, Liu et al.~\cite{2019-NQE} find that the performance of NCS regresses with shorter queries. To address this issue, they explore an additional way of using neural networks in code search. They develop NQE, a neural model that takes in a set of keywords and predicts a set of keywords to expand the query to NCS. NQE with NCS can perform better than using NCS alone. Specifically, NQE is an encoder-decoder model, given a query as input, which outputs the most likely sets of expanded keywords. NQE learns to predict keywords that co-occur with the query keywords in the underlying corpus, which helps productively expand the query. Besides, beam search is also utilized to obtain the top-k most likely sequences of method names. This enhances the performance of NQE in finding the most relevant keywords for expanding the query.

Huang et al.~\cite{2019-QESC} find a potential issue where expanded queries may inadvertently include irrelevant terms. This phenomenon, known as the ``overexpansion problem'', can lead to confusion within the search engine and subsequently result in worse outcomes. To avoid the overexpansion problem, they propose a novel query expansion algorithm based on the semantics of change sequences, named QESC. Change sequences are generated from the commits of each method. They contain the changed terms (the new or the deleted code terms) as well as dependent terms (the unchanged code terms). As we mentioned earlier, DBN is trained on the change sequences. Then, QESC uses the DBN to generate changed terms which are composed of operation, label, and role. If the operation of the term is new, it means the term is relevant to the original query, promoting QESC to expand the query with its label. According to the obtained changed terms, QESC guarantees the expanded terms are relevant to the original query, and second-pass retrieval with an expanded query will perform better. The same approach has been applied to the query expansion technique with code changes (QECC)~\cite{2018-Query-Expansion-Based-on-Code-Changes} to improve the ability to infer the expansion words. Similar to QESC, QECC also extracts (changes, contexts) pairs from the abstract syntax trees (ASTs) of changed methods, which are used to detect changes and extract contexts.
However, the difference between the two lies in the output of the inference model, the association-based inference model trained by QECC can directly infer the suggested words to expand the query. Precisely, upon receiving a query, QECC retrieves initial results from the code corpus. Subsequently, the model extracts the contexts of these initial results and deduces potential expansion words by assessing the basis of initial results. Ultimately, the model constructs an expanded query by incorporating these expansion words into the original query. This expanded query is then utilized to execute a final search within the code corpus.

Questions and Answers (Q\&A) from programming forums that contain abundant exchanging knowledge about programming issues are crucial resources for code retrieval and annotation. However, Hu et al.~\cite{2020-Unsupervised-Software-Repositories} find that mining software repositories in such open and unrestricted forums is challenging, since the posts can be arbitrary and noisy. To overcome this challenge, they propose Code-Description Mining Framework (CodeMF), an unsupervised framework to eliminate noisy posts and extract high-quality software repositories from programming forums. Specifically, CodeMF is the combination of two proposed frameworks kernel principal component analysis (KPCA) and wavelet time-frequency transform feature fusion (WTFF) which are applied in extracting high-quality software repositories. KPCA is the principal component analysis tool used to reduce the dimension of features collected from the software repositories, and then further extract high-quality mappings between query-code pairs. Regarding the novel framework WTFF, it transforms the multiple dimension features into a time-frequency domain through wavelet transformation to reduce the computational complexity. This helps to extract the principal components of the software repository features more easily. Finally, CodeMF leverages the QECK~\cite{2016-Query-Expansion-Based-on-Crowd-Knowledge-for-CS} to retrieve the code snippets. As we introduced above, QECK uses the Q\&A context-text pairs to expand the NL queries. CodeMF enhances the Q\&A Pairs Search Engine which is a crucial component of QECK. Thus, it offers high-quality expanded keywords, leading to improved code search performance. 

Cai et al.~\cite{2021-Search-for-Compatible-Code} find the retrieved source code from the existing code search techniques is not compatible with local programming language since the evolution and production of multiple versions of libraries. To solve this issue, they propose DCSE, a deep code search model based on evolving information. Specifically, DCSE first deeply excavates evolved code tokens and evolution descriptions in the code evolution process. 
It then treats evolved code tokens and evolution descriptions as one feature of source code and code descriptions, respectively. Therefore, DCSE can retrieve the source code that is compatible with the local programming language. DCSE embeds source code and its code descriptions into a high-dimensional shared vector space. It retrieves the initial result using the given query from the repositories. If the initial search result is incompatible with the local programming language, users could add the error report of IDE to query for a second-time search. This is due to the error report being semantically related to the evolution description while evolution description is one feature of code description, so the distance between the expanded query vector and the compatible source code vector will be closer. Finally, DCSE can leverage the expanded query to retrieve the compatible source code.

Query reformulation is a widely utilized technology that can be regarded as similar to query expansion for enriching user requirements and enhancing the outcomes of code search. However, Mao et al.~\cite{2023-Self-Supervised-Query-Reformulation} find that training a query reformulation model requires a large parallel corpus of query pairs (i.e., the original query and a reformulated query) that are confidential and not publicly available. This restricts the practicality of query reformulation in software development processes. Therefore, they propose SSQR, a self-supervised query reformulation method that does not rely on any parallel query corpus. 
Specifically, SSQR treats query reformulation as a masked language modeling task conducted on an extensive unannotated corpus of queries. SSQR extends T5~\cite{2020-T5} (a sequence-to-sequence model-based on Transformer) with a new pre-training objective named corrupted query completion (CQC), which randomly masks words within a complete query and trains T5 to predict the masked content. Subsequently, for a given query to be reformulated, SSQR identifies potential locations for expansion and leverages the pre-trained T5 model to generate the appropriate content to fill these gaps. In this way, SSQR enhances the performance of code search from the unsupervised query reformulation.

\subsubsection{Query Transformation}
\label{subsubsec:query_transformation}
Considering the gap in syntax and structure between the query in natural language and the code snippet in programming language, there are naturally some obstacles if directly matched between them. To address this issue, some researchers believe that finding an intermediate pattern to bridge this gap holds promise as a viable solution. They successively propose some query transformation techniques, which can convert natural language queries into other forms or augment with those forms, such as Q\&A posts and other customized forms~\cite{2018-Augmenting-and-Structuring-User-Query, 2022-Enriching-Query-Semantics-for-CS-with-Reinforcement-Learning}. In short, query transformation turns the direct matching problem between query and code into an indirect one. From the third line of Figure~\ref{fig:evolution_query_feature_mining}, it is observed that the research enthusiasm for both query transformation and query reduction is comparable but notably lower than that observed for query expansion. The subsequent sections will delve into a detailed discussion of how these works conduct the process of query transformation.

Sirres et al.~\cite{2018-Augmenting-and-Structuring-User-Query} find a significant issue where source code terms such as method names and variable types are often different from conceptual words mentioned in a query. This is called a mismatch problem which occurs from the poorly documented or non-explicit names of source code. To reduce this mismatch problem, they present COCABU, a novel approach that leverages common developer questions and the associated expert answers to transform the augmented queries from the original user queries according to the relevant Q\&A sites.  Specifically, in the first step of COCABU, the search proxy takes an original query as input and returns a set of relevant posts collected from developer Q\&A sites as an output. The obtained Q\&A posts are used to find out how natural language concepts can be translated into program elements, that is, to collect potential translation rules. These translation rules can alleviate the vocabulary mismatch problem between user queries and source code. 
To transform the original query into the augmented query, the code query generator module extracts structural code entities from code snippets in Q\&A sites. 
Besides, the code query generator only considers the accepted answers in Q\&A sites. Thereafter, based on the Lucene search engine, COCABU preserves the terms from search results and combines them with the types of structural code entities collected from Q\&A sites (e.g., unqualified/partially qualified method invocations or classes) to form the augmented query. 
Finally, the code search engine takes an augmented query produced by the code query generator and provides a list of search results to the developer. 

Ling et al.~\cite{2021-DGMS} find most of code search techniques ignore the deep structured features when processing both queries and code snippets. To address this problem, they propose an end-to-end deep graph matching and searching (DGMS) model based on graph neural networks for the task of semantic code retrieval. 
Considering the rich, important semantic structure information within the queries and code snippets, DGMS builds the graph-structured data to represent that structure information.  
Specifically, for the natural language query text, DGMS builds the text graph based on the constituency parse tree~\cite{2019-Introduction-to-natural-language-processing} and word ordering features. These features provide both constituent and ordering information of sentences to establish the graph-structured data. In a nutshell, DGMS performs query transformation by converting the query text into a text graph. For code snippets, DGMS also generates corresponding code graphs. 
After transforming both queries and code snippets into unified graph-structured data, DGMS uses the proposed graph matching and searching model to retrieve the best matching code snippet.

Wang et al.~\cite{2022-Enriching-Query-Semantics-for-CS-with-Reinforcement-Learning} find that the user query is relatively shorter than the code description (also known as code comments) and limited in context. It implies a knowledge gap between the query and the code description. The code description contains more semantic keywords like code snippets rather than the query. However, existing research ignores this gap, resulting in low code search accuracy for code search models trained based on code descriptions rather than real queries. 
To reduce the impact of the knowledge gap, Wang et al. propose a query-enriched code search model called QueCos. 
QueCos performs query transformation by generating the corresponding code descriptions from the given query. Those descriptions are utilized for improving the code search performance. 
Specifically, QueCos collects the code-description pairs dataset from the GitHub. The designed crawler saves the code snippets referred in the Stack Overflow posts and the corresponding code descriptions. 
Then, a query semantic enriching model is designed to generate the corresponding descriptions for queries based on the collected dataset, during which reinforcement learning is adopted to enable the code snippets retrieved by the generated descriptions to be ranked higher. 
The generated descriptions are treated as semantically enriched queries and not necessarily to be exactly close to the ground-truth descriptions. 
Finally, both the semantically enriched queries and original queries are employed for the ultimate code search.

Existing code search techniques have primarily relied on intricate matching and attention-based mechanisms. However, Tang et al.~\cite{2023-Hyperbolic-Code-Retrieval} find that those techniques often lead to computational and memory inefficiencies, posing a significant challenge to their real-world applicability. To tackle this challenge, they propose a novel technique, the Hyperbolic Code QA Matching (HyCoQA). HyCoQA leverages the unique properties of Hyperbolic space to express connections between code snippets and their corresponding queries. Specifically, HyCoQA transforms the code search task into a Q\&A pair matching paradigm. It constructs a dataset with triple matches characterized as '$\langle$negative code, description, positive code$\rangle$'. In this case, the primary objective is to maximize the margin between the scores of the correct Q\&A pair and the negative Q\&A pair, ensuring that the retrieve system can robustly differentiate between accurate and inaccurate solutions based on the given description. Thus, these triple matches are subsequently processed via a static BERT embedding layer, yielding initial embeddings. A novel mathematical concept, hyperbolic geometry is used by HyCoQA. Unlike traditional Euclidean spaces, hyperbolic spaces excel at depicting hierarchical structures, which often underlie the relationship between code and its corresponding natural language description. Therefore, HyCoQA utilizes a hyperbolic embedder to transform the initial embeddings into hyperbolic space, calculating distances between the codes and descriptions. The process concludes by implementing a scoring layer based on these distances and leveraging hinge loss for model training.

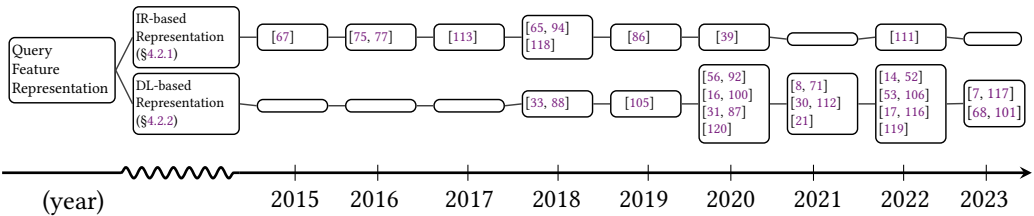
\begin{figure*}[!th]
    \centering
    \resizebox{0.98\textwidth}{!}{
        \begin{forest}
            for tree={
                grow=east,
                reversed=true,
                anchor=base west,
                parent anchor=east,
                child anchor=west,
                font=\large,
                rectangle,
                draw=black,
                rounded corners,
                base=left, 
                align=left,
                minimum width=4em,
                edge+={darkgray, line width=1pt},
                s sep=3pt,
                inner xsep=2pt,
                inner ysep=3pt,
                line width=1pt,
                ver/.style={rotate=90, child anchor=north, parent anchor=south, anchor=center},
            },
            where level=1{text width=7em, font=\normalsize}{},
            where level=2{text width=4.5em, font=\normalsize}{},
            where level=3{text width=4.5em, font=\normalsize}{},
            where level=4{text width=4.5em, font=\normalsize}{},
            where level=5{text width=4.5em, font=\normalsize}{},
            where level=6{text width=4.5em, font=\normalsize}{},
            where level=7{text width=4.5em, font=\normalsize}{},
            where level=8{text width=4.5em, font=\normalsize}{},
            where level=9{text width=4.5em, font=\normalsize}{},
            [Query\\Feature\\Representation, text width=7em
                [IR-based\\Representation\\(\S\ref{subsubsec:IR-based_feature_representation})
                    [\hspace{7.5pt}\cite{2015-CodeHow}
                        [\cite{2016-Query-Expansion-Based-on-Crowd-Knowledge-for-CS, 2016-Learning-to-rank-code-examples-for-CS}
                            [\hspace{7.5pt}\cite{2017-IECS}
                                [\cite{2018-Interactive-Query-Reformulation, 2018-Augmenting-and-Structuring-User-Query}\\
                                \cite{2018-Expanding-Queries-for-Code-Search}
                                    [\hspace{9.5pt}\cite{2019-Automatic-Query-Reformulation-for-CS}
                                        [\hspace{9.5pt}\cite{2020-Unsupervised-Software-Repositories}
                                            [[\hspace{7.5pt}\cite{2022-Learning-deep-semantic-model-for-CS}
                                            []]]
                                        ]
                                    ]
                                ]
                            ]
                        ]
                    ]
                ]
                [DL-based\\Representation\\(\S\ref{subsubsec:DL-based_feature_representation})
                    [
                        [
                            [
                                [\cite{2018-Retrieval-on-Source-Code-A-Neural-CS, 2018-Deep-Code-Search}
                                    [\hspace{7.5pt}\cite{2019-Multi-modal-Attention-for-Code-Rerieval}
                                        [\cite{2020-adaptive-deep-code-search, 2020-CARLCS}\\
                                            \cite{2020-pscs, 2020-CoNCRA}\\
                                            \cite{2020-CRaDLe, 2020-csda}\\
                                            \cite{2020-OCoR}
                                            [\cite{2021-Search-for-Compatible-Code, 2021-At-CodeSM}\\
                                                \cite{2021-Multimodal-Representation-for-Neural-Code-Search, 2021-TabCS}\\
                                                \cite{2021-SAN-CS}
                                                [\cite{2022-CodeHunter, 2022-CSRS}\\
                                                \cite{2022-Enriching-Query-Semantics-for-CS-with-Reinforcement-Learning, 2022-JessCS}\\
                                                \cite{2022-fined-grained-co-attentive-representation-learning, 2022-Incorporating-Code-Structure-and-Quality-in-Deep-Code-Search}\\
                                                \cite{2022-EAGCS}
                                                    [\cite{2023-CSSAM, 2023-deGraphCS}\\
                                                    \cite{2023-Hyperbolic-Code-Retrieval, 2023-Self-Supervised-Query-Reformulation}]
                                                ]
                                            ]
                                        ]
                                    ]
                                ]
                            ]
                        ]
                    ]
                ]
            ]
        \end{forest}
    }

    \begin{tikzpicture}[x=30]
        \draw[line width=1pt] (0,0) -- (1.5,0);
        \draw[snake=snake, line width=1pt, segment amplitude=2pt, segment length=6pt] (1.5,0) -- (3,0);
        \draw[-stealth, line width=1pt,] (3,0) -- (13,0);
        
        \foreach \x in {3.9,5,6.2,7.4,8.6,9.7,10.8,12,13.1}
        \draw (\x cm,3pt) -- (\x cm,-3pt);
        
        \draw (0.9,0) node[below=3pt] {(year)} node[above=3pt] {};
        \draw (1.8,0) node[below=3pt] {} node[above=3pt] {};
        \draw (2.8,0) node[below=3pt] {} node[above=3pt] {};
        \draw (3.7,0) node[below=3pt] {2015} node[above=3pt] {};
        \draw (4.7,0) node[below=3pt] {2016} node[above=3pt] {};
        \draw (5.9,0) node[below=3pt] {2017} node[above=3pt] {};
        \draw (7.0,0) node[below=3pt] {2018} node[above=3pt] {};
        \draw (8.1,0) node[below=3pt] {2019} node[above=3pt] {};
        \draw (9.2,0) node[below=3pt] {2020} node[above=3pt] {};
        \draw (10.3,0) node[below=3pt] {2021} node[above=3pt] {};
        \draw (11.4,0) node[below=3pt] {2022} node[above=3pt] {};
        \draw (12.4,0) node[below=3pt] {2023} node[above=3pt] {};
        
    \end{tikzpicture}

    \caption{Evolution of feature representation techniques in query end}
    \label{fig:evolution_query_feature_representation}
\end{figure*}

\subsection{Query Feature Representation}
\label{subsec:query_feature_representation}

\subsubsection{IR-based Feature Representation}
\label{subsubsec:IR-based_feature_representation}
Information retrieval (IR) is the technique of finding relevant information from large amounts of information according to the needs of users~\cite{2001-Modern-information-retrieval}.   
In the field of code retrieval (i.e., code search), the IR technique has naturally been introduced by researchers. There are many mature IR-based methods to handle natural language queries. Those methods represent queries as vectors or plain texts, and researchers utilize these representations to optimize the query end, thereby enhancing the final results of code search. 
Generally speaking, traditional IR methods represent queries as some form of index (such as keyword terms and vectors), which can reflect the semantic information contained within queries. Thereby, the code search model can utilize these representations to optimize the query end and improve the final performance. 
The first line of Figure~\ref{fig:evolution_query_feature_representation} showcases the code search works that adopt IR-based techniques to represent query features. It is observed that most early code search efforts employ IR techniques to represent queries. 
This demonstrates that IR technology is simple and effective in feature representation, and we will describe these works in detail later.

Plain text is a common feature representation method in the field of information retrieval, which refers to preprocessed text from the query. The plain text representation can be used to optimize the query end. As mentioned in Section~\ref{subsec:query_feature_mining}, CodeHow~\cite{2015-CodeHow} expands the query with the APIs and performs code retrieval. To find the related APIs, CodeHow needs to represent the descriptions of AIPs and the given query before calculating their similarity. 
Therefore, Codehow utilizes plain text as the representation of features used in traditional IR methods, while preprocessing of the text is required before generating the plain text representation.
In the preprocessing stage, CodeHow finds some words that appear very often and do not have a definite meaning, called stop words (e.g., on, the, are, etc.). To reduce the impact of those meaningless words in the queries and descriptions, they remove those stop words after adopting the text normalization. Thereafter, they also perform stemming. The goal of those preprocessing steps is to represent the text into keyword terms that can be used to compute the similarity by VSM.
In the code retrieve stage, the method of representing the queries expanded with APIs is also similar to the preprocessing stage. The expanded queries, represented in the form of plain text, are used for BM to match with code snippets.

Nie et al.~\cite{2016-Query-Expansion-Based-on-Crowd-Knowledge-for-CS} also employ plain text to represent query features and query expansion word features. As mentioned in Section~\ref{subsubsec:query_expansion}, they propose a query expansion tool called QECK. In QECK, there are two times of feature representation. The first one involves representing the original query for first-pass retrieval, while the second one involves representing the expanded candidate words to select the most suitable ones. These two instances of feature representation both utilize plain text. 
In this way, QECK directly processes the text into the search without converting it into vectors. Specifically, the original text is split by Camel-case and separators (e.g., `\_'). Then, it filters these words by removing the stop words. Besides, the remaining words are handled by stemming. After the preprocessing, the obtained plain text can be used to retrieve the code snippets. Lu et al.~\cite{2018-Interactive-Query-Reformulation} and Hu et al.~\cite{2020-Unsupervised-Software-Repositories} also utilize a similar feature representation rule to extract the plain text representation for user queries. They regard the processed plain text representation as a keyword set that can retrieve the code more easily. 

In order to extract the plain text representation of queries more accurately, Rahman et al.~\cite{2019-Automatic-Query-Reformulation-for-CS} propose a Part-of-Speech (POS) tagging on the query before normal preprocessing. POS can extract meaningful words such as nouns and verbs from the query. Once a query is submitted, they first perform POS and then apply standard natural language preprocessing (i.e., stop word removal, splitting, and stemming) on the query to extract the stemmed words. This more precise query feature representation can further enhance the final performance of code retrieval.

Using traditional IR methods to represent queries, whether expanded or reduced, as vectors is also a common approach. For example, the use of VSM as a method for representing queries as vectors has been applied in many techniques. For example, Sirres et al.~\cite{2018-Augmenting-and-Structuring-User-Query} and Niu et al.~\cite{2016-Learning-to-rank-code-examples-for-CS} use the VSM to determine the relevancy of the user query. It represents the query as a vector where the term is computed by the TF-IDF weighting. 

Another widely used IR method is the bag-of-words model (BOW)~\cite{2013-Efficient-Estimation-of-Word-Representations-in-Vector-Space}. 
BOW represents a text document as a collection of vocabulary, disregarding the sequence and context of words, only focusing on the frequency of word occurrences within the document. For example, Wu et al.~\cite{2022-Learning-deep-semantic-model-for-CS} employ the BOW to represent each term in a normalized query and generate the term vector to represent the semantic feature. 
In BOW, text is regarded as an unordered collection of vocabulary, disregarding its grammar and context. Feature information of query is obtained using the TF-IDF, which calculates the term frequency (TF) and inverse document frequency (IDF). 
Thereafter, the query vectors are calculated by summing up the vectors of each term. Yang et al.~\cite{2017-IECS} also adopt BOW to represent the query feature. After the normal preprocessing, they convert the query and the intents into a BOW model. The output from the BOW model is regarded as the representation of the query. To map queries with identifiers that are used to expand the queries, Zhang et al.~\cite{2018-Expanding-Queries-for-Code-Search} utilize the continues bag-of-words model (CBOW) to convert the original query into a vector representation.
The continuous bag-of-words model is an improved version of the traditional BOW model. Unlike the BOW model, which focuses solely on the frequency of words within a document, the CBOW model pays attention to the vocabulary information in the context. Therefore, the semantic features represented by the query vectors generated through CBOW are more accurate.

\subsubsection{DL-based Feature Representation}
\label{subsubsec:DL-based_feature_representation}
DL-based feature representation applies deep neural networks to encode the given feature data to produce semantic-preserving numerical vector representations (also known as embeddings). It has also been widely used to optimize query feature representation in code search. When code search techniques apply DL techniques/models to transform queries into embeddings, models will extract fine-grained semantic and structural information in queries. 
The second line of Figure~\ref{fig:evolution_query_feature_representation} presents the code search techniques that apply DL techniques to represent queries. These techniques transfer different neural networks/models from the natural language processing (NLP) or computer vision (CV) fields to encode queries. We will introduce these techniques in subsequent paragraphs.

Ling et al.~\cite{2020-adaptive-deep-code-search} and Kong et al.~\cite{2022-CodeHunter} utilize FastText~\cite{2017-FastText} to build word vector representations. 
FastText is an open-source model released by Facebook. Researchers can directly download pre-trained FastText models from open-source repositories to represent the queries they need to process.
Besides, FastText employs CBOW and Skip-gram models to learn word vector representations. It focuses on the context of words, generating dense, low-dimensional vectors for each word.
Therefore, word representations learned by FastText are more reliable and effective.

Sachdev et al.~\cite{2018-Retrieval-on-Source-Code-A-Neural-CS} are among the first researchers to adopt DL techniques to represent query features. They propose a neural code search tool named NCS.
In NCS, query feature representation is decomposed into two steps: 1) building word embeddings and 2) building document embeddings. NCS treats a query as a document. 
In step 1), NCS uses a variant of the Word2vec model, called FastText~\cite{2017-FastText} to build word vector representations (i.e., embeddings). It employs the continuous skip-gram model with a window size of 5, i.e., all pairs of words within a distance of 5 are considered nearby words. 
In step 2), NCS expresses the intent of the query in the same high–dimensional vector space as the word embeddings, by aggregating the representations of all the words extracted from the query. 
The authors found that building document embeddings by simply averaging word embeddings does not work well for their purposes. Therefore, they further tried three variants of the combination method: i. Average over all the words;
ii. Average over the unique words in each document;
iii. Weighted average of all unique words in a document according to the following equations.
\begin{gather}
    v_d = u\left(\sum_{w \in d}{u(v_w) \cdot \text{tfidf}(w, d, C)}\right)
    \\
    \text{tfidf}(w, d, C) = \frac{1 + \log\text{tf}(w, d)}{\log|C|/\text{df}(w, C)}
    \label{equ:document_embedding}
\end{gather}
where $d$ is a multiset of words representing a document; $C$ is the corpus containing all documents; $u$ is a normalizing function where $u(v) = \frac{v}{|v|}$; \text{tfidf}, short for term frequency–inverse document frequency, is a function that assigns a weight for a given word in a given document~\cite{2017-Neural-Information-Retrieval}. A word has a higher weight if it appears frequently in the document but is also penalized if it appears in too many documents in the corpus. Their experiments show that the weighted average method works significantly better than the others.

Recurrent Neural Network (RNN) is also widely used to encode queries. For example, Gu et al.~\cite{2018-Deep-Code-Search} utilize an RNN to represent query features. It is known that RNN has a recurrent structure within the network where hidden layers are recurrently used for computation. Therefore, unlike traditional feed-forward neural networks, RNN can encode sequential queries using its internal memory. The hidden state represents the feature of the query as the final output of RNN. The query representation through RNN can be computed as:
\begin{gather}
    h_i = \tanh\left(W\left[h_{i-1}, w_i\right]\right)
    \label{equ:rnn}
\end{gather}
where each hidden state $h_i$ is generated from the previous hidden state $h_{i-1}$ and $w_i$ is the one-hot representation of each word in query; while $W$ is the matrix of trainable parameters in the RNN, while $\tanh(\cdot)$ is a non-linearity activation function of the RNN.

Wan et al.~\cite{2019-Multi-modal-Attention-for-Code-Rerieval} propose a comprehensive multi-model representation method for source code called MMAN. To match with the source code representation, MMAN exploits the standard LSTM to learn the representation of the given query. 
In MMAN, building a query feature representation consists of two steps: 1) embedding the query into a vector and 2) generating the hidden state from LSTM.
In step 1), MMAN builds a word embedding layer that uses the one-hot embedding function to compute the given query into embedding.
In step 2), MMAN applies an LSTM to represent the query. As an improved version of RNN, LSTM can better capture long-term dependencies. Each LSTM unit contains an input gate, a memory cell, and an output gate. The unit receives the embedding from step 1) and generates the hidden state given to the next unit. The last hidden state can be used as the query feature representation denoted $h_i^{query}$. It can be computed as:
\begin{gather}
    h_i^{query} = \mathbf{LSTM}\left(h_{i-1}^{query}, w(q_i)\right)
    \label{equ:lstm}
\end{gather}
where 
$w(\cdot)$ is the word embedding layer to embed each word into a vector; $q$ is the input of the word embedding layer which donates the query; the last hidden state $h_{d}^{query}$ is the final output of the LSTM. Due to its outstanding performance and efficiency, LSTM is also widely employed by other code search techniques for query feature representation. For instance, CSSAM~\cite{2023-CSSAM} and DCSE~\cite{2021-Search-for-Compatible-Code} all follow almost the same process as MMAN.

Unlike NCS in building document embeddings step, CARLCS-CNN~\cite{2020-CARLCS}, CoNCRA~\cite{2020-CoNCRA}, and CSRS~\cite{2022-CSRS} introduce the use of the convolutional neural network (CNN) for query feature representation extraction after obtaining the word embedding. As for the query, it is usually short, but it contains informative keywords that reflect the intention of the user. Therefore, CNN is suitable for extracting query feature representation from word embedding. The feature representation $D$ of a query can be computed as: 
\begin{gather}
    v_i = f\left(W_i \ast E_{i:i+h-1} + b\right)
    \\
    \mathrm{D}_{\mathrm{h}}=\left[v_{1}, v_{2}, \ldots, v_{d}\right]
    \\
    \mathrm{D} = \mathrm{D}_{\mathrm{1}} \oplus \ldots \oplus \mathrm{D}_{\mathrm{n}}
\end{gather}
where $E$ is the embedding of the query; while $W$ is the convolution kernels for convolution operation; and $f$ is a non-linear function such as the hyperbolic tangent. After generating the feature scores $v$, the embedding matrix $D_h$ can also be obtained. Thereafter, the final feature representation $D$ is accomplished by merging all the embedding matrices.

Differing slightly from the aforementioned methods, EAGCS~\cite{2022-EAGCS} and CRaDLe~\cite{2020-CRaDLe} incorporate a maxpooling layer after LSTM. They take the hidden states generated by each LSTM unit and feed them into the maxpooling layer to obtain the final query representation. The purpose of adding a maxpooling layer is to perform further feature extraction, obtaining the significant features. Therefore, the final representation of the query, denoted $\bm{v}^{query}$, can be formulated as:
\begin{gather}
    \bm{v}^{query} = \mathbf{maxpooling}\left(\left[h_1,\dots, h_i\right]\right) 
    \label{equ:maxpooling}
\end{gather}
where $\mathbf{maxpooling}(\cdot)$ refers to maxpooling layer; the $h$ is the hidden state from each LSTM unit and calculated by Equation~(\ref{equ:lstm}).

As a variant of the traditional LSTM, bi-directional LSTM (Bi-LSTM) is also employed by Ren et al.~\cite{2020-csda}, Wang et al.~\cite{2022-Enriching-Query-Semantics-for-CS-with-Reinforcement-Learning}, and Meng~\cite{2021-At-CodeSM} to generate query feature representations. Unlike the traditional LSTM, Bi-LSTM simultaneously considers the past and future information at each time step in the sequence to better capture contextual relationships. Bi-LSTM consists of two LSTM layers, one processing the input sequence from left to right in terms of time steps (i.e., forward LSTM), and the other processing the input sequence from right to left (i.e., backward LSTM). The outputs of these two LSTM layers are concatenated to form a representation that incorporates bidirectional contextual information. The final feature representation generated by Bi-LSTM for the query can be formalized as:
\begin{gather}
    \bm{v}^{query} = h_{q} = \left[\overrightarrow{h}_{q}, \overleftarrow{h}_{q}\right] 
    \label{equ:Bi-LSTM}
\end{gather}
where $\overrightarrow{h}_{q}$ and $\overleftarrow{h}_{q}$ are the forward and backward hidden states produced by the final layer of LSTM.

Considering the self-attention mechanism is suitable for capturing the structural and semantic features within query sequences, Fang et al.~\cite{2021-SAN-CS}, Zhu et al.~\cite{2020-OCoR}, Gu et al.~\cite{2021-Multimodal-Representation-for-Neural-Code-Search}, Kong et al.~\cite{2022-JessCS}, and Sun et al.~\cite{2020-pscs} utilize the self-attention mechanism to extract features and generate representations.
The framework they designed for constructing the query feature representation by the self-attention mechanism can be divided into two steps: 1) generating word embedding of query and 2) obtaining feature representation from the self-attention mechanism. 
In step 1), the framework applies a word embedding tool to build the embedding layer. Sun et al. and Kong et al. select the one-hot code to represent the word embedding of the query. The rest of the techniques exploit the well-known word embedding tool, called word2vec~\footnote{\url{https://github.com/danielfrg/word2vec/}}. Word2vec uses the CBOW to embed words. Compared with one-hot representation, distributed representation produced by word2vec can build semantic relations between different words. 
In step 2), the word embedding generated by the embedding layer is further embedded by the self-attention mechanism. The self-attention mechanism can effectively extract features from the query and generate the feature representations. 
The following equation simply summarizes the above process of generating a query feature representation through the embedding layer and attention mechanism.
\begin{gather}
    \bm{v}^{query} = \mathbf{Att}\left(E\left(s^{query}\right)\right) 
    \label{equ:att}
\end{gather}
where $s^{query}$ is the given query treated as a list of words; while $E(\cdot)$ represents a word embedding layer; $\bm{v}^{query}$ is the final output of attention mechanism.

To fully understand the relationships between query and code, Deng et al.~\cite{2022-fined-grained-co-attentive-representation-learning} propose a code search model named FcarCS. 
The approach of representing query features in FcarCS is similar to the aforementioned self-attention framework, but it replaces the self-attention mechanism with the co-attention mechanism as the feature-extracted tool. 
FcarCS constructs a new fine-grained co-attention mechanism to learn interdependent representations for each code and query, respectively. The workflow of this co-attention mechanism can be divided into computing semantic association, extracting semantic information, and calculating semantic vector parts. According to those steps, the co-attention mechanism establishes dynamic attention relationships between queries and code snippets. Therefore, it can explore more fine-grained semantic correlations between each code snippet and query, and enrich the query feature representations.

There also exists an architecture that utilizes both attention mechanisms and a co-attention mechanism to generate query feature representations. 
To bridge the semantic gap between code snippet and query effectively and efficiently, Xu et al.~\cite{2021-TabCS} propose a two-stage attention-based model for code search, called TabCS. The first stage leverages attention mechanisms to extract semantics from queries considering their textual features. The second stage leverages a co-attention mechanism to capture the semantic correlation between queries and code snippets. Therefore, the co-attention mechanism contributes to better query representation. 

Zeng et al.~\cite{2023-deGraphCS} also propose an embedding module similar to MD-JEnn, but they replace Bi-LSTM with LSTM. They utilize the structure of an LSTM layer along with an attention layer to accurately extract feature representations containing keyword information from user queries. Besides, Hu et al.~\cite{2020-NJACS} design a similar structure to obtain feature representation. However, instead of using one-hot representation or pre-trained word2vec embeddings to generate the word embedding, they use a structure embeddings matrix to incorporate word-level structure information and get the structure embeddings of queries. 

Yu et al.~\cite{2022-Incorporating-Code-Structure-and-Quality-in-Deep-Code-Search} propose a novel deep neural network named Method-Description-Joint Embedding Neural Network (MD-JEnn), which uses a joint embedding technique to model the semantic relation between code snippets and descriptions. The description embedding module (DE-Module) is a component of MD-JEnn that embeds natural language descriptions (queries) into vectors. 
The word embedding model and Bi-LSTM are used in MD-JEnn to embed queries into query vectors first.
Since some words in a description are important, it is necessary to assign higher weights to these important words. Thus, MD-JEnn introduces an attention mechanism to aggregate the query vectors of the description into a feature-represented vector by calculating a scalar weight for each vector of the description word. The individual vectors are aggregated to a feature-represented vector $\bm{v}$ via attention:
\begin{gather}
    \small
    a_i = \frac{\exp{({\widetilde{h_{i}}}^\intercal \cdot \alpha)}}{\sum_{j=1}^{n}\exp{({\widetilde{h_{j}}}^\intercal \cdot \alpha)}}
\end{gather}
\begin{gather}
    \small
    \bm{v} = \sum_{i=1}^{n}\alpha_{i} \cdot \widetilde{h}_{i}
\end{gather}
where $h$ is the the hidden states of the previous Bi-LSTM layer, which represents the query vector and $\alpha_i$ is a learnable matrix initialized in random, $\alpha_i$ also represents the attention weight of each $\widetilde{h}_{i}$.

Tang et al.~\cite{2023-Hyperbolic-Code-Retrieval} leverage the unique properties of Hyperbolic space to express the feature of user queries. Unlike traditional Euclidean spaces, hyperbolic spaces excel at depicting hierarchical structures. Therefore, they first exploit the BERT~\cite{2019-BERT} (a pre-trained NLP model) as the embedding layer to embed the original query. Thereafter, they use an advanced hyperbolic embedder to encode the query embedding into hyperbolic spaces as the final representation. This representation can express the feature of the query deeply and easily to match with the code snippets.

SSQR~\cite{2023-Self-Supervised-Query-Reformulation} utilizes masked language modeling task (MLM) to conduct the extensive unannotated corpus of queries to reformulate queries. 
SSQR takes the T5~\cite{2020-T5} (a pre-trained NLP model) as the backbone model for the MLM task since it has a sequence-to-sequence architecture.
The queries are converted into the embeddings by the MLM and masked out a span of 15\% consecutive words from a randomly selected position.
The purpose of the MLM task is to encourage the model to learn contextual relationships and better understand the semantics of language by inferring missing words in a given context. Following MLM training, the T5 model can accurately capture the semantic meanings of each word in the queries and generate high-quality query embeddings. 
Therefore, these query embeddings are used for the final code retrieval.

\finding{1}{
Existing code search techniques mainly optimize the query end from two aspects: query feature mining and query feature representation. 
For query feature mining methods, query reduction, query expansion, and query transformation continue to receive attention. Among them, query expansion overall is the most, followed by query transformation, and query reduction. Compared with query reduction, research on query transformation has become more popular in recent years and continues to be a hot trend. 
For query feature representation methods, most early code search techniques apply information retrieval techniques (e.g., plain text, bag of words, and TF-IDF) to represent query features. In recent years, with the emergence of deep learning technology, most researchers have turned to using neural networks (e.g., FastText, CNN, RNN, LSTM, Bi-LSTM, BERT, and T5) and attention mechanisms (e.g., self-attention and co-attention mechanisms) to represent query features, and this phenomenon will obviously continue.
}

\section{Answering RQ2: What are the code-end optimization methods in code search studies?}
\label{sec:Answering_RQ2}
Understanding code semantics also plays an important role in code search. Only by correctly understanding what a code snippet is doing can we better match it to the query. As a human-readable text written in a specific programming language, code snippet contains semantic information not only in the text but also in its structure, such as AST. To learn about the semantics of the code snippet more comprehensively, it is often necessary to leverage such semantic information and convert it into other representations that can be used to match queries at a later stage. As a result, many code-end optimization techniques have been proposed to capture and understand the semantics of the code snippet to improve the effectiveness and efficiency of code search. 

As illustrated in Figure~\ref{fig:framework_of_code_search}, similar to the query end, the optimization techniques devised for the code end can likewise be divided into two primary parts: code feature mining and code feature representation. For code feature mining, existing works mainly focus on mining textual and structural features of the code snippet. The textual features include method names, API sequences, and tokens. The structural features, also referred to as intermediate representations, encompass abstract syntax tree (AST), data flow graph (DFG), control flow graph (CFG), program dependence graph (PDG), and variable-based flow graph (VFG). Details of the above textual and structural features are discussed in Section~\ref{subsec:code_feature_mining}.
For code feature representation, similar to the query end, the techniques used to represent code features can also be divided into two major classes, i.e., IR-based feature representation and DL-based feature representation. Details of these code feature representation methods are discussed in Section~\ref{subsec:code_feature_representation}. In the following subsections, we will discuss these code features and their representation ways in detail, extract their commonalities, find their differences, and summarize the development trends.

\begin{figure*}[!th]
    \centering
    \resizebox{0.98\textwidth}{!}{
        \begin{forest}
            for tree={
                grow=east,
                reversed=true,
                anchor=base west,
                parent anchor=east,
                child anchor=west,
                base=center,
                font=\large,
                rectangle,
                draw=black,
                rounded corners,
                align=left,
                minimum width=4em,
                edge+={darkgray, line width=1pt},
                s sep=3pt,
                inner xsep=2pt,
                inner ysep=3pt,
                line width=1pt,
                ver/.style={rotate=90, child anchor=north, parent anchor=south, anchor=center},
            },
            where level=1{text width=10em, font=\normalsize,}{},
            where level=2{text width=4.5em, font=\normalsize,}{},
            where level=3{text width=4.5em, font=\normalsize,}{},
            where level=4{text width=4.5em, font=\normalsize,}{},
            where level=5{text width=4.5em, font=\normalsize,}{},
            where level=6{text width=4.5em, font=\normalsize,}{},
            where level=7{text width=4.5em, font=\normalsize,}{},
            where level=8{text width=4.5em, font=\normalsize,}{},
            where level=9{text width=4.5em, font=\normalsize,}{},
            [Code \\ Feature \\ Mining
                [Code textual \\ 
                    feature mining (\S\ref{subsec:code_textual_feature_mining})
                    [method \\ name
                        [
                            [\cite{2018-Deep-Code-Search, 2018-Retrieval-on-Source-Code-A-Neural-CS}
                                [\hspace{3.5pt} \cite{2019-Multi-modal-Attention-for-Code-Rerieval}
                                    [\cite{2020-CARLCS, 2020-csda}      [\cite{2021-TabCS, 2021-At-CodeSM} \\ \cite{2021-Search-for-Compatible-Code, 2021-CodeMatcher} \\ \cite{2021-SAN-CS}
                                            [\cite{2022-Incorporating-Code-Structure-and-Quality-in-Deep-Code-Search, 2022-CSRS} \\ \cite{2022-fined-grained-co-attentive-representation-learning, 2022-Enriching-Query-Semantics-for-CS-with-Reinforcement-Learning}
                                            \\ \cite{2022-JessCS, 2022-CodeHunter}[]
                    ]]]]]]]
                    [API \\ sequence 
                        [
                            [\hspace{7.5pt}\cite{2018-Deep-Code-Search}
                                [\hspace{7.5pt}\cite{2019-Automatic-Query-Reformulation-for-CS}
                                    [\hspace{7.5pt}\cite{2020-CARLCS}
                                        [\cite{2021-TabCS, 2021-MuCoS} \\ \cite{2021-SAN-CS, 2021-Search-for-Compatible-Code}
                                            [\cite{2022-fined-grained-co-attentive-representation-learning, 2022-CodeHunter} \\ \cite{2022-CSRS, 2022-JessCS}[]
                    ]]]]]]]
                    [tokens
                        [\hspace{7.5pt}\cite{2015-CodeHow}
                            [\cite{2018-Deep-Code-Search, 2018-a-neural-framework} \\ \cite{2018-Expanding-Queries-for-Code-Search, 2018-Augmenting-and-Structuring-User-Query}
                                [\cite{2019-Multi-modal-Attention-for-Code-Rerieval, 2019-Coacor}
                                    [\cite{2020-CARLCS, 2020-adaptive-deep-code-search} \\ \cite{2020-MP-CAT-model, 2020-csda} \\ \cite{2020-CoNCRA, 2020-COSEA} \\ \cite{2020-OCoR, 2020-CodeBERT}
                                        [\cite{2021-Multimodal-Representation-for-Neural-Code-Search, 2021-At-CodeSM} \\ \cite{2021-CodeMatcher, 2021-Search-for-Compatible-Code} \\ \cite{2021-TabCS, 2021-SAN-CS}
                                            [\cite{2022-Incorporating-Code-Structure-and-Quality-in-Deep-Code-Search, 2022-Learning-deep-semantic-model-for-CS} \\ \cite{2022-CSRS, 2022-fined-grained-co-attentive-representation-learning} \\ \cite{2022-JessCS, 2022-Accelerating-Code-Search} \\ \cite{2022-Enriching-Query-Semantics-for-CS-with-Reinforcement-Learning, 2022-SPT-Code}
                                            [\cite{2023-CSSAM, 2023-Hyperbolic-Code-Retrieval}]
                    ]]]]]]]
                ]
                [Code structural \\ feature mining (\S\ref{subsubsec:code_structural_feature_mining})
                    [AST
                        [
                            [
                                [\hspace{6.5pt}\cite{2019-Multi-modal-Attention-for-Code-Rerieval}
                                    [\cite{2020-MP-CAT-model, 2020-CRaDLe} \\ \cite{2020-NJACS, 2020-pscs}
                                        [\cite{2021-TabCS, 2021-Multimodal-Representation-for-Neural-Code-Search} \\ \cite{2021-At-CodeSM}
                                            [\hspace{7.5pt}\cite{2022-SPT-Code}[]
                    ]]]]]]]
                    [DFG
                        [
                            [
                                [\hspace{6.5pt}\cite{2019-Multi-modal-Attention-for-Code-Rerieval} 
                                    [\hspace{7.5pt}\cite{2020-CRaDLe}
                                        [\hspace{7.5pt}\cite{2021-GraphCodeBERT}
                                            [
                                                [\hspace{9.5pt}\cite{2023-CSSAM}]
                    ]]]]]]]
                    [CFG
                        [
                            [
                                [\hspace{6.5pt}\cite{2019-Multi-modal-Attention-for-Code-Rerieval} 
                                    [\hspace{7.5pt}\cite{2020-CRaDLe}[[[]
                    ]]]]]]]
                    [PDG
                        [
                            [
                                [
                                    [\hspace{7.5pt}\cite{2020-CRaDLe}
                                        [
                                            [\hspace{6.5pt}\cite{2022-EAGCS}[]
                    ]]]]]]]
                    [others
                        [
                            [
                                [
                                    [
                                        [\hspace{7.5pt}\cite{2021-DGMS}
                                            [\hspace{6.5pt}\cite{2022-Incorporating-Code-Structure-and-Quality-in-Deep-Code-Search}
                                                [\cite{2023-deGraphCS, 2023-GraphSearchNet}]
                    ]]]]]]]
                ]
            ]
        \end{forest}
    }

    \begin{tikzpicture}[x=30]
        \draw[line width=1pt] (0,0) -- (1.5,0);
        \draw[snake=snake, line width=1pt, segment amplitude=2pt, segment length=6pt] (1.5,0) -- (3,0);
        \draw[line width=1pt] (3,0) -- (5.5,0);
        \draw[snake=snake, line width=1pt, segment amplitude=2pt, segment length=6pt] (5.5,0) -- (6.2,0);
        \draw[-stealth, line width=1pt,] (6.2,0) -- (13,0);
        
        \foreach \x in {5.5, 6.8, 8, 9.3, 10.6, 11.9, 13.1}
        \draw (\x cm,3pt) -- (\x cm,-3pt);
        
        \draw (0.9,0) node[below=3pt] {(year)} node[above=3pt] {};
        \draw (1.8,0) node[below=3pt] {} node[above=3pt] {};
        \draw (2.8,0) node[below=3pt] {} node[above=3pt] {};
        \draw (5.2,0) node[below=3pt] {2015} node[above=3pt] {};
        \draw (6.4,0) node[below=3pt] {2018} node[above=3pt] {};
        \draw (7.6,0) node[below=3pt] {2019} node[above=3pt] {};
        \draw (8.8,0) node[below=3pt] {2020} node[above=3pt] {};
        \draw (10,0) node[below=3pt] {2021} node[above=3pt] {};
        \draw (11.3,0) node[below=3pt] {2022} node[above=3pt] {};
        \draw (12.5,0) node[below=3pt] {2023} node[above=3pt] {};
        
    \end{tikzpicture}

    \caption{Evolution of feature mining techniques in code end}
    \label{fig:evolution_code_feature_mining_new}
\end{figure*}
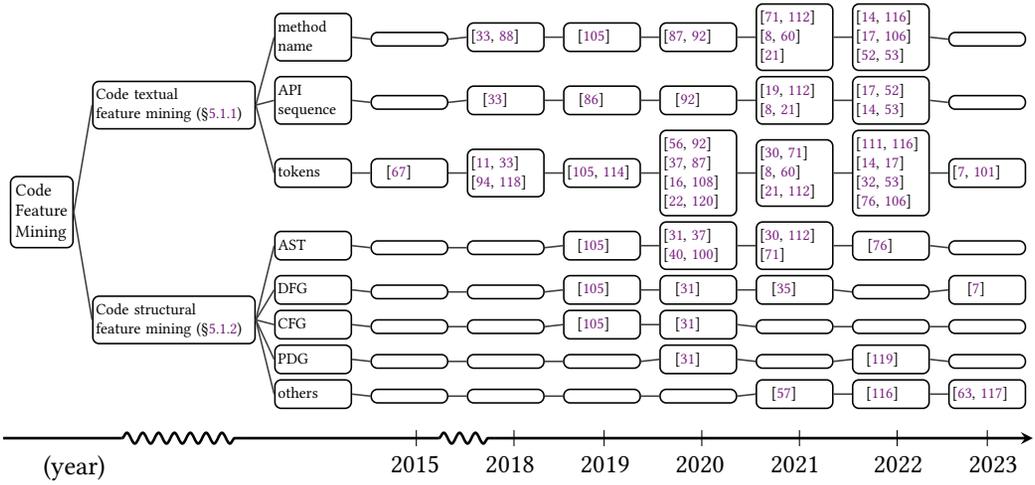

\subsection{Code Feature Mining}
\label{subsec:code_feature_mining}

\subsubsection{Code Textual Feature Mining}
\label{subsec:code_textual_feature_mining}
\noindent\newline
The code snippets we discuss in this survey are at the method (function) level. A method-level code snippet usually consists of a method name, parameters (optional) and a method body. In code search, commonly used code textual features include method name, API sequence, and tokens, all of which contain semantic information~\cite{2022-JessCS}.

\textbf{Method name}. 
The method name often outlines the functionality (semantics) of a code snippet (at the method level)~\cite{2020-CARLCS}. 
Therefore, it plays an important role in understanding the code snippet. 
As shown in the first line of Figure~\ref{fig:evolution_code_feature_mining_new}, many works~\cite{2018-Deep-Code-Search, 2018-Retrieval-on-Source-Code-A-Neural-CS, 2019-Multi-modal-Attention-for-Code-Rerieval, 2020-CARLCS, 2020-csda, 2021-TabCS, 2021-CodeMatcher, 2021-At-CodeSM, 2021-Search-for-Compatible-Code, 2021-SAN-CS, 2022-Incorporating-Code-Structure-and-Quality-in-Deep-Code-Search,2022-CodeHunter, 2022-JessCS, 2022-fined-grained-co-attentive-representation-learning, 2022-CSRS, 2022-Enriching-Query-Semantics-for-CS-with-Reinforcement-Learning} mine and utilize this feature. The extraction of method names is relatively simple and intuitive, and some works describe this process in detail. 
For example, CodeMatcher~\cite{2021-CodeMatcher} develops a tool called JAnalyzer~\footnote{\url{https://github.com/liuchaoss/janalyzer}}, which transforms a method into an AST with the Javaparser~\footnote{\url{https://github.com/javaparser/javaparser}} library and then extracts method name by traversing the AST. CodeHunter~\cite{2022-CodeHunter} first uses JDT (i.e., Eclipse Java Development Tools) to construct the AST of the source code, and then invokes the $\mathtt{getName()}$ method in the MethodDeclaration class to obtain the method name. Considering that a method name typically consists of multiple words, it is a common practice to split the camel–case (camelCase) or snake–case (snake\_case) concatenated method names into multiple separate tokens for later embedding.

\textbf{API sequence}. 
API (Application Programming Interface) is of great significance in code search. It refers to a predefined and encapsulated function. Rather than implementing a method from scratch, programmers often invocate APIs in their code to facilitate their development activities~\cite{2016-DeepAPI}. What’s more, the naming of an API usually briefly describes its functionality (semantics), so the API sequence extracted from a code snippet can help code search models understand how this code snippet is implemented. 
As illustrated in the second line of Figure~\ref{fig:evolution_code_feature_mining_new}, many works~\cite{2018-Deep-Code-Search, 2019-Automatic-Query-Reformulation-for-CS, 2020-CARLCS, 2021-TabCS, 2021-Search-for-Compatible-Code, 2021-SAN-CS, 2021-MuCoS, 2022-CodeHunter, 2022-JessCS, 2022-fined-grained-co-attentive-representation-learning, 2022-CSRS} make use of this feature. Next, we will introduce two ways to extract API sequences. Rahman et al.~\cite{2019-Automatic-Query-Reformulation-for-CS} extract API sequences from Stack Overflow posts with island parsing techniques. They first isolate code snippets from the HTML source of each answer from Stack Overflow using <$code$> tags, and then use a regular expression for Java class to extract the API class tokens having camel case notation. DeepCS~\cite{2018-Deep-Code-Search}, CodeHunter~\cite{2022-CodeHunter}, and DCSE~\cite{2021-Search-for-Compatible-Code} follow the method described in DeepAPI~\cite{2016-DeepAPI} to extract an API sequence from each Java method.
They all employ the Eclipse JDT compiler to parse and traverse the AST. 
After obtaining the AST, the API sequence is generated by the following rules~\cite{2018-Deep-Code-Search, 2022-JessCS}:
\begin{itemize}
    \item For a constructor invocation $\mathtt{new~C()}$, they produce $\mathtt{C.new}$.
    
    \item For a method call $\mathtt{o.m()}$ where $\mathtt{o}$ is an instance of class $\mathtt{C}$, they create $\mathtt{C.m}$.
    
    \item For a method call passed as a parameter, they append the method before the calling method.

    \item For a statement sequence $s_1; s_2; \dots; s_N$, they extract the API sequence $a_i$ from each statement $s_i$ and concatenate them to form the API sequence $a_1\text{-}a_2\text{-}\dots\text{-}a_N$.
    
    \item For conditional statements such as $if(s_1)\{s_2;\}else\{s_3;\}$, they produce a sequence from all possible branches, i.e.,$a_1\text{-}a_2\text{-}a_3$, where $a_i$ is the API sequence extracted from the statement $s_i$.
    
    \item For loop statements such as $while(s_1)\{s_2;\}$, they generate a sequence $a_1\text{-}a_2$, where $a_1$ and $a_2$ are API sequences extracted from statement $s_1$ and $s_2$, respectively.
\end{itemize}

\textbf{Tokens}. 
Tokens are bags of words that are parsed from the method body of a code snippet. They include useful information such as constant names, variable names, and comments written by the developer.
Usually, data preprocessing for tokens removes duplicate words, stop words, and keywords in the programming language, which improves the quality of tokens.
As a result, as presented in the third line of Figure~\ref{fig:evolution_code_feature_mining_new}, almost all code search techniques~\cite{2015-CodeHow, 2018-Deep-Code-Search, 2018-a-neural-framework, 2018-Augmenting-and-Structuring-User-Query, 2018-Expanding-Queries-for-Code-Search, 2019-Multi-modal-Attention-for-Code-Rerieval, 2019-Coacor, 2020-CARLCS, 2020-csda, 2020-COSEA, 2020-CoNCRA, 2020-MP-CAT-model, 2020-adaptive-deep-code-search, 2020-CodeBERT, 2020-OCoR, 2021-Search-for-Compatible-Code, 2021-TabCS, 2021-Multimodal-Representation-for-Neural-Code-Search, 2021-SAN-CS, 2021-At-CodeSM, 2021-CodeMatcher, 2022-Incorporating-Code-Structure-and-Quality-in-Deep-Code-Search, 2022-Learning-deep-semantic-model-for-CS, 2023-CSSAM, 2022-JessCS, 2022-fined-grained-co-attentive-representation-learning, 2022-CSRS, 2022-SPT-Code, 2022-Enriching-Query-Semantics-for-CS-with-Reinforcement-Learning, 2022-Accelerating-Code-Search, 2023-Hyperbolic-Code-Retrieval} take advantage of this feature. Most of them tokenize code snippets through camel case splitting or snake case splitting. Zhang et al.~\cite{2018-Expanding-Queries-for-Code-Search} focus on query-end optimization and aim to extend natural-language queries with API class-names, so they keep them intact.

\subsubsection{Code Structural Feature Mining}
\label{subsubsec:code_structural_feature_mining}
\noindent\newline
In code search, commonly used code structural features include AST, DFG, CFG, PDG, and other structural features.

\textbf{AST}.
Abstract syntax tree (AST) abstractly represents the syntactic structure of a code snippet in the form of a tree~\cite{2019-Multi-modal-Attention-for-Code-Rerieval}. Each node in the tree denotes a structure in the code snippet, such as loop structure, conditional judgment structure, method call, and variable declaration~\cite{2021-TabCS}. 
There is no doubt that AST is very helpful in understanding the code as these structures denote the logic of the code. Consequently, AST is the most commonly used structural feature in code search techniques~\cite{2019-Multi-modal-Attention-for-Code-Rerieval, 2020-CRaDLe, 2020-pscs, 2020-NJACS, 2020-MP-CAT-model, 2021-TabCS, 2021-Multimodal-Representation-for-Neural-Code-Search, 2021-At-CodeSM, 2022-SPT-Code}. They use various different AST parsing tools to generate AST. For instance, MMAN~\cite{2019-Multi-modal-Attention-for-Code-Rerieval} parses C code into AST via an open source tool named Clang~\footnote{\url{http://clang.llvm.org/}}. PSCS~\cite{2020-pscs} extracts AST paths using PathMiner~\cite{2019-PathMiner}, an open source java library for mining path-based representations of code. At-CodeSM~\cite{2021-At-CodeSM} employs javalang to generate an AST from the code snippet. SPT-Code~\cite{2022-SPT-Code} first uses an AST parser~\footnote{\url{https://tree-sitter.github.io/tree-sitter/}} to get the AST. Then, it utilizes a simplified version
of structure-based traversal (SBT)~\cite{2018-Deep-Code-Comment-Generation} called XML-like SBT (X-SBT) to traverse the AST and parse it into a sequence. By employing SBT, the resulting sequence can be reduced by more than half in length.

\textbf{DFG}. 
Data flow graph (DFG) is a type of intermediate representation. 
DFG refers to the data flow of a program, which describes how the data in a piece of code flows and how it is processed. Many code search works~\cite{2019-Multi-modal-Attention-for-Code-Rerieval, 2020-CRaDLe, 2021-GraphCodeBERT, 2023-CSSAM} use DFG as a feature to capture the data dependencies between code elements. For instance, Hu et al.~\cite{2023-CSSAM} establish a code semantic representation graph (CSRG) based on AST and DFG, which is a graph structure more compact than AST. GraphCodeBERT~\cite{2021-GraphCodeBERT} is a pre-trained model for programming language, which utilizes data flow in the pre-training stage. It first parses a code $C$ into an AST, whose terminals (leaves) are used to identify the variable sequence, denoted as $V = \{v_1, v_2, \dots, v_k\}$. Then, it takes each variable as a node of the graph, and a direct edge $\epsilon = \langle v_i, v_j \rangle$ from $v_i$ to $v_j$ means that the value of $j$-th variable comes from $i$-th variable. Taking $x = expr$ as an example, $expr$ is an expression (e.g., $(a+b)*c$), edges from all variables in $expr$ to $x$ are added into the graph. The set of directed edges is denoted as $E = \{\epsilon_1, \epsilon_2, \dots, \epsilon_l\}$ and the graph $\mathcal{G}(C) = (V, E)$ is data flow used to represent dependency relation between variables of the code.

\textbf{CFG}.
Control flow graph (CFG) is another type of intermediate representation of the code feature. CFG means the computation and control flow of a program, which has the function of representing all possible execution paths for the program. 
CFG is also a code feature commonly used by code search techniques~\cite{2019-Multi-modal-Attention-for-Code-Rerieval, 2020-CRaDLe} to capture control dependencies between code elements. For example, Wan et al.~\cite{2019-Multi-modal-Attention-for-Code-Rerieval} directly make use of the CFG of a code snippet in their code search model. They first parse C functions into CFGs via an open-source tool named SVF~\cite{2016-SVF}~\footnote{\url{https://github.com/SVF-tools/SVF}}. Then, for nodes with the same statement, they keep the nodes that appear first in the output of SVF, remove their child nodes, and connect the children of their child nodes to them. For nodes without statements, they delete them and link their child nodes to their parent nodes. 

\textbf{PDG}. Program dependence graph (PDG) can represent the data dependencies and control dependencies of each operation in a program~\cite{1987-PDG}. It is built on AST, but not as deep as AST in structure, and reserves only the execution paths that will affect the execution results~\cite{2020-CRaDLe}. Gu et al.~\cite{2020-CRaDLe} utilize PDG to extract code structures. They propose CRaDLe, a code retrieval model based on semantic dependency learning, to learn the matching relationship between the code and description pairs which helps to retrieve the related code snippets. Zhao et al.~\cite{2022-EAGCS} establish a statement-level advanced program dependence graph (APDG), which introduces the statement execution information and control logic missed in PDG. Based on APDG, they propose EAGCS, a novel code search approach that largely enhances the expression of structural and semantic information in source code. APDG helps EAGCS to learn a deeper understanding of code vectors which improves the performance of retrieving code snippets from given query. 

\textbf{Others}. Some code search techniques define new structural features based on those above. 
For instance, a variable-based flow graph (VFG) is proposed by Zeng et al. in their work DeGraphCS~\cite{2023-deGraphCS}. They think that tokens and structural features cannot accurately express the in-depth semantics of source code. To overcome this limitation, they propose VFG, which integrates tokens, data flow, and control flow. 
VFG is constructed on LLVM IR instructions. Specifically, to construct VFG, they first extract the identifiers in each LLVM IR instruction as nodes. Then, they build data dependencies and control dependencies between nodes according to different types of instructions. The data dependencies are built based on the address operation instructions (e.g., ``load'') and the computation-related constructions (e.g., ``add''). The control dependencies are constructed based on the jump instructions (e.g., ``br'') and address operation instructions.
Finally, they apply an optimization mechanism to remove the redundant nodes without changing the semantic information. 

DGMS~\cite{2021-DGMS} and GraphSearchNet~\cite{2023-GraphSearchNet} both construct a program graph to represent the source code, which helps them precisely learn unified semantic relation representation of the source code and queries. Given a code snippet $c$, a program graph is a multi-edged directed graph $g(V,\bm{A})\in\mathcal{G}$ extracted from $c$. $V$ is a set of nodes built on the AST and $\bm{A}\in\{0, 1\}^{k \times m \times m}$ is the adjacency matrix representing the relationships between the nodes (i.e., the edges), where $k$ and $m$ are the total number of edge types and nodes in $g$, respectively. Particularly, the leaf nodes of AST correspond to the identifier in the code snippet, and the non-leaf nodes represent different compilation units such as ``Assign'', ``BinOp'', ``Expr''. In addition, GraphSearchNet also builds the syntactic edges (i.e., ``NextToken'', ``SubToken'') and data-flow edges (i.e., ``ComputedFrom'', ``LastUse'', ``LastWrite'') based on AST nodes, which can represent the code snippet.

Noticing that the API sequence generated by traversing the AST tree ignores semantics contained in the structure of the code snippet, SQ-DeepCS~\cite{2022-Incorporating-Code-Structure-and-Quality-in-Deep-Code-Search} introduces a novel method called \emph{program slice} to preserve structural information. To generate a program slice, first, the AST of a code snippet is parsed, and then different processing methods are applied to different statements extracted from the code snippet. Taking the loop statement as an example, the program slice reserves its judgment conditions and loop body and adds the $for$ keyword. For instance, $for(c1; c2; c3)\{s4; \}$ is converted to $for(p2)\{p4; \}$, where $p2$ and $p4$ are the program slices of condition $c2$ and statement $s4$.

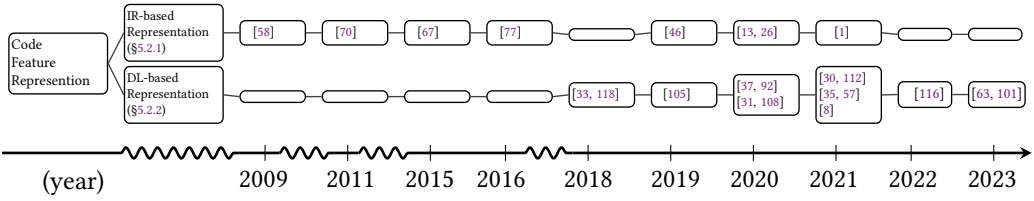
\begin{figure*}[!th]
    \centering
    \resizebox{0.98\textwidth}{!}{
        \begin{forest}
            for tree={
                grow=east,
                reversed=true,
                anchor=base west,
                parent anchor=east,
                child anchor=west,
                base=left,
                font=\large,
                rectangle,
                draw=black,
                rounded corners,
                align=left,
                minimum width=4em,
                edge+={darkgray, line width=1pt},
                s sep=3pt,
                inner xsep=2pt,
                inner ysep=3pt,
                line width=1pt,
                ver/.style={rotate=90, child anchor=north, parent anchor=south, anchor=center},
            },
            where level=1{text width=7em, font=\normalsize,}{},
            where level=2{text width=4.5em, font=\normalsize,}{},
            where level=3{text width=4.5em, font=\normalsize,}{},
            where level=4{text width=4.5em, font=\normalsize,}{},
            where level=5{text width=4.5em, font=\normalsize,}{},
            where level=6{text width=4.5em, font=\normalsize,}{},
            where level=7{text width=4.5em, font=\normalsize,}{},
            where level=8{text width=4.5em, font=\normalsize,}{},
            where level=9{text width=4.5em, font=\normalsize,}{},
            [Code \\ Feature \\ Represention, text width=7em
                [IR-based \\ Representation \\ (\S\ref{subsubsec:IR-based_code_feature_representation})
                    [\hspace{7.5pt}\cite{2009-Sourcerer}
                    [\hspace{7.5pt}\cite{2011-Portfolio}
                    [\hspace{7.5pt}\cite{2015-CodeHow}
                    [\hspace{7.5pt}\cite{2016-Learning-to-rank-code-examples-for-CS}
                    [[\hspace{7.5pt}\cite{2019-ROSF}
                    [\cite{2020-Enhancing-Example-based-Code-Search, 2020-recommendation-based-on-java-code-analysis-and-search}
                    [\hspace{11.5pt}\cite{2021-FACER}[[]]]]]]]]]
                    ]
                ]
                [DL-based \\ Representation \\ (\S \ref{subsec:feature-level_code-end-optimization})
                    [[[[[\cite{2018-Deep-Code-Search, 2018-Expanding-Queries-for-Code-Search}
                    [\hspace{6.5pt}\cite{2019-Multi-modal-Attention-for-Code-Rerieval}
                    [\cite{2020-CARLCS, 2020-MP-CAT-model}\\ \cite{2020-CRaDLe, 2020-COSEA}
                    [\cite{2021-TabCS, 2021-Multimodal-Representation-for-Neural-Code-Search}\\ \cite{2021-DGMS, 2021-GraphCodeBERT}\\ \cite{2021-Search-for-Compatible-Code}
                    [\hspace{6.5pt}\cite{2022-Incorporating-Code-Structure-and-Quality-in-Deep-Code-Search}
                    [\cite{2023-Hyperbolic-Code-Retrieval, 2023-GraphSearchNet}]]]]]]]]]]
                ]
            ]
        \end{forest}
    }

    \begin{tikzpicture}[x=30]
        \draw[line width=1pt] (0,0) -- (1.5,0);
        \draw[snake=snake, line width=1pt, segment amplitude=2pt, segment length=6pt] (1.5,0) -- (3,0);
        \draw[line width=1pt] (3,0) -- (3.5,0);
        \draw[snake=snake, line width=1pt, segment amplitude=2pt, segment length=6pt] (3.5,0) -- (4.2,0);
        \draw[line width=1pt] (4.2,0) -- (4.5,0);
        \draw[snake=snake, line width=1pt, segment amplitude=2pt, segment length=6pt] (4.5,0) -- (5.2,0);
        \draw[line width=1pt] (5.2,0) -- (6.6,0);
        \draw[snake=snake, line width=1pt, segment amplitude=2pt, segment length=6pt] (6.6,0) -- (7.2,0);
        \draw[-stealth, line width=1pt,] (7.2,0) -- (13,0);
        \foreach \x in {3.5,4.6,5.7,6.7,7.8,8.9,10,11.1,12.1,13.2}
        \draw (\x cm,3pt) -- (\x cm,-3pt);
        
        \draw (0.9,0) node[below=3pt] {(year)} node[above=3pt] {};
        \draw (1.8,0) node[below=3pt] {} node[above=3pt] {};
        \draw (2.8,0) node[below=3pt] {} node[above=3pt] {};
        \draw (3.3,0) node[below=3pt] {2009} node[above=3pt] {};
        \draw (4.4,0) node[below=3pt] {2011} node[above=3pt] {};
        \draw (5.4,0) node[below=3pt] {2015} node[above=3pt] {};
        \draw (6.3,0) node[below=3pt] {2016} node[above=3pt] {};
        \draw (7.4,0) node[below=3pt] {2018} node[above=3pt] {};
        \draw (8.5,0) node[below=3pt] {2019} node[above=3pt] {};
        \draw (9.5,0) node[below=3pt] {2020} node[above=3pt] {};
        \draw (10.5,0) node[below=3pt] {2021} node[above=3pt] {};
        \draw (11.5,0) node[below=3pt] {2022} node[above=3pt] {};
        \draw (12.5,0) node[below=3pt] {2023} node[above=3pt] {};
        
    \end{tikzpicture}

    \caption{Evolution of feature representation techniques in code end}
    \label{fig:evolution_code_feature_representation}
\end{figure*}

\subsection{Code Feature Representation}
\label{subsec:code_feature_representation}

\subsubsection{IR-based Feature Representation}
\label{subsubsec:IR-based_code_feature_representation}
\noindent\newline
As mentioned earlier, code Search is a successful application of IR where the information retrieved is code snippets. 
The core idea of IR-based code search methods is to treat source code as text, so the code and the query can be matched by their textual similarity. The first line of Figure~\ref{fig:evolution_code_feature_representation} presents the code search works that adopt IR-based techniques to represent code features. 
We will describe these studies in detail later.

While an approach based entirely on IR was effective in the early days, as time goes by, researchers have come to realize that treating code as just plain text does not provide a comprehensive understanding of its semantics. As a result, later techniques combine it with some additional information, such as the relational representations of code~\cite{2009-Sourcerer}, API~\cite{2015-CodeHow, 2021-FACER}, comment~\cite{2020-recommendation-based-on-java-code-analysis-and-search} and keywords~\cite{2020-Enhancing-Example-based-Code-Search}, etc. Another research~\cite{2011-Portfolio} aims at making it easier for developers to make use of the code snippet retrieved by showing other possibly related functions and the call graph between them. In this section, we will introduce several typical IR-based code feature representation techniques.

In most applications of mining and searching software corpora, the code’s structural aspects, as well as the relevant metadata surrounding it, are always ignored. To utilize this information, Linstead~\cite{2009-Sourcerer} et al. propose Sourcer, which combines standard text IR techniques with source-specific heuristics and a relational representation of code. It employs a relational database consisting of two tables to store the data: (1) program entities: uniquely identifiable elements from the source code; and (2) their relations: any dependency between two entities. In addition, they also store compact representations of attributes for fast retrieval of search results, including keywords from FQNs and comments, and fingerprints used to support structural searches of source code. All entity keywords and metadata are indexed using Lucene to support fast search. In this way, Sourcerer provides a comprehensive, multi-modal platform for searching and finding reusable software components.

Niu et al.~\cite{2016-Learning-to-rank-code-examples-for-CS} propose a code search approach that utilizes a machine learning technique to automatically train a ranking schema. They represent each code snippet as a 12-dimension vector, each dimension denotes the value of a feature extracted from the code snippet. The 12 features are classified into four categories: similarity, popularity, code metrics, and context. The similarity is the textual similarity between a query and candidate code snippets, computed by the Vector Space Model (VSM)~\cite{1975-VSM}. The popularity category has two features: frequency and probability. Frequency is the number of times that the frequent method call sequence of a candidate code snippet occurs in the corpus, while probability is the likelihood of following the method call sequence in a candidate code example. The code metrics category contains eight features that only reflect characteristics of a code snippet regardless of the query, including the line length, the average number of identifiers per line, and so on. The context similarity refers to the similarity between the context of the query and the candidate code snippets.

Similar to Niu et al.~\cite{2016-Learning-to-rank-code-examples-for-CS}, Jiang et al.~\cite{2019-ROSF} also leverages textual and structural features like the textual similarity between a query and a candidate code snippet and line length of a code snippet to represent the code. Besides, they also consider the topic similarity between a query and a code snippet since a code snippet can be considered as a textual document
describing one or more technical topics. They first generate a term-by-document matrix $M$, given a collection of code snippets and queries. Then, they employ Latent Dirichlet Allocation (LDA)~\cite{2003-LDA} to identify the latent variables (topics) hidden in the data and generate as output a topic-by-document matrix. A generic entry $\theta_{ij}$ of this matrix denotes the probability of the $j^{th}$ document to belong to the $i^{th}$ topic. Finally, the topic similarity between a query and a code snippet can be calculated based on this matrix.

\subsubsection{DL-based Feature Representation}
\label{subsec:feature-level_code-end-optimization}
\noindent\newline
DL-based feature representation on the code-end optimizes the capture and representation of code features (semantics) by introducing advanced deep learning techniques, thereby improving the performance of code search. In this technique, both queries and code snippets are transformed to feature vector representations (also called embeddings). Embedding learns to represent entities (e.g., words, sentences, and graphs) as vectors with the aim of making vector representations of similar entities close to each other~\cite{ 2013-Efficient-Estimation-of-Word-Representations-in-Vector-Space, 2013-Distributed-Representations-of-Words-and-Phrases-and-their-Compositionality}. These embeddings are randomly initialized and then fixed via an end-to-end supervised training paradigm. 
The purpose of embedding training in code search is to bridge the lexical gap between code snippets in programming languages and queries in natural language, so as to better understand the semantics of the code snippets. 
In the task of code search, both code and query are embedded into a unified vector space through Joint Embedding, which is a technique to jointly embed heterogeneous data. As a result, similar concepts with different modalities are close to each other in this space and the code relevant to a query can be measured by the distance between their vector representations, such as their cosine similarity~\cite{2018-Deep-Code-Search}. 

In 2018, Gu et al.~\cite{2018-Deep-Code-Search} first introduced deep learning to the field of code search. They propose a novel deep neural network called CODEnn to learn a unified vector representation of both code snippets and natural language queries. Based on CODEnn, they develope a prototype named DeepCS to support code retrieval. 
After DeepCS, many DL-based feature representation techniques have sprung up. In this section, we will first discuss how they leverage advanced deep learning techniques to represent a single feature of a code snippet, and how they integrate these features to build high-performance code search models.

To begin with, we will introduce the representation methods for the textual features of a code snippet as mentioned in Section~\ref{subsec:code_textual_feature_mining}.

\textbf{For the method name}, DeepCS~\cite{2018-Deep-Code-Search} and SQ-DeepCS~\cite{2022-Incorporating-Code-Structure-and-Quality-in-Deep-Code-Search} both use a Recurrent Neural Network (RNN) to encode the sequence of the tokens obtained by splitting the method name, as RNN can capture the semantics of sequence information well. It is computed using the same formula~(\ref{equ:rnn}) mentioned in Section~\ref{subsubsec:DL-based_feature_representation}. DCSE~\cite{2021-Search-for-Compatible-Code} employs Bi-LSTM to learn the semantic information since the method name has word order. 
Empirically finding that the average length of each method name sequence is only 2 or 3 in its training data, CARLCS-CNN~\cite{2020-CARLCS} applies a Convolutional Neural Networks (CNN) instead of RNN, which is supposed to be good at extracting robust and abstract features. 
Therefore, in CARLCS-CNN, the representation of the method name is calculated as:
\begin{gather}
    \boldsymbol{m}_{1: n}=\boldsymbol{m}_{1} \oplus \boldsymbol{m}_{2} \oplus \dots \oplus \boldsymbol{m}_{n}
    \\
    \boldsymbol{c}_{i} = f\left(W_{M} * \boldsymbol{m}_{i: i+h-1} + \boldsymbol{b}\right)
    \\
    M_h=[\bm{c}_1, \bm{c}_2, \dots, \bm{c}_{n-h+1}]
\end{gather}
where $\boldsymbol{m}_i \in \mathbb{R}^k$ is the $k$-dimensional word vector; $\oplus$ is the concatenation operator; $W_M \in \mathbb{R}^{k \times h}$ is a \textit{filter} involved in the convolution operation, which is applied to a window of $h$ words to produce a feature; $\boldsymbol{b} \in \mathbb{R}$ is a bias term; $*$ is the convolution operator and $f$ is a non-linear function such as the hyperbolic tangent; $M_h$ is a \textit{feature map} produced after applying the filter to each possible window of words in the method name. What's more, CARLCS-CNN uses three types of filters with varying window sizes $h$ from 2 to 4, with the number of each type of filter set to $d$. Then, it completes the convolution operation through these filters to extract three distinctive feature maps, i.e., $M_{h_1}, M_{h_2}, M_{h_3} \in \mathbb{R}^{d \times (n-h+1)}$, respectively, and finally concatenates them into a feature matrix $M$:
\begin{gather}
    M=M_{h_1} \oplus M_{h_2} \oplus M_{h_3}
\end{gather}

TabCS~\cite{2021-TabCS} utilizes attention mechanism-based search models to improve the efficiency of training and testing. Let $\boldsymbol{m}_i \in \mathbb{R}^k$ be a $k$-dimensional word initial vector corresponding to the $i$-th word in a method name. Given a sequence of length n \{$\boldsymbol{m}_1, \dots, \boldsymbol{m}_n$\}, the attention weight for each $\boldsymbol{m}_i$ is computed as follows:
\begin{gather}
    \alpha_{m_i}=\frac{exp(\boldsymbol{a}_{m_i}\cdot\boldsymbol{m}_i^T)}{\sum_{i=1}^nexp(\boldsymbol{a}_{m_i}\cdot\boldsymbol{m}_i^T)}
    \label{equ:attention}
\end{gather}
where the attention vector $\boldsymbol{a}_{m_i}$ is optimized during model training. The attention weight for each initial vector is computed by applying the softmax function to the product of the initial vectors and attention vectors. Then, it multiplies each initial vector with its corresponding attention weight, and concatenates the resulting weighted vectors:
\begin{gather}
    M=\alpha_{m_1}\boldsymbol{a}_1 \oplus \alpha_{m_2}\boldsymbol{a}_2 \oplus \dots \oplus \alpha_{m_n}\boldsymbol{a}_n
    \label{equ:attention_concatenation}
\end{gather}

\textbf{For the API sequence}, similar to the handling of the method name, DeepCS~\cite{2018-Deep-Code-Search} encodes it into a vector representation using an RNN with maxpooling. In view of the dynamic sequential features of the API sequence, CARLCS-CNN~\cite{2020-CARLCS} implements Bi-LSTM to do the embedding, using the formula~(\ref{equ:Bi-LSTM}) mentioned in Section~\ref{subsubsec:DL-based_feature_representation}. Then, the API sequence is encoded by concatenating all the output hidden states to a feature matrix. Finally, TabCS~\cite{2021-TabCS} performs an attention mechanism on the randomly initialized feature matrix of the API sequence, the same as its process to the method name and tokens.

\textbf{For the tokens}, considering that tokens are the informative keywords of code, CARLCS-CNN~\cite{2020-CARLCS} uses CNN to encode them. COSEA~\cite{2020-COSEA} also leverages CNN, since CNN has the natural ability to capture locality information which can be used for capturing code blocks' information. Moreover, COSEA utilizes layer-wise attention to learn the code semantic representation. After transforming a code snippet into embedding vectors, the model can get longer and longer code block representation through convolutional modules. Finally,  attentive pooling is performed on these block representations and the ultimate semantic embedding of the code snippet is obtained.

DeepCS~\cite{2018-Deep-Code-Search} and SQ-DeepCS~\cite{2022-Incorporating-Code-Structure-and-Quality-in-Deep-Code-Search} simply encode tokens via a multilayer perceptron (MLP), i.e., the conventional fully connected layer, as tokens are unordered in the source code. Let $\boldsymbol{\tau}_i \in \mathbb{R}^d$ represent the embedded representation of the token $\tau_i$, $\boldsymbol{W}^\Gamma$ represent the matrix of trainable parameters in the MLP, the embedding vector $\boldsymbol{h}_i$ of the $i$-th token is computed as:
\begin{gather}
    \label{equ:mlp}
    \boldsymbol{h}_i = \tanh(\boldsymbol{W}^\Gamma\boldsymbol{\tau}_i), \forall i = 1, 2, \dots, N_\Gamma
\end{gather}
Next, the individual vectors are summarized to a single vector $t$ through maxpooling:
\begin{gather}
    \boldsymbol{t} = \text{maxpooling}([\boldsymbol{h}_1, \dots, \boldsymbol{h}_{N_\Gamma}])
\end{gather}

Zhang et al.~\cite{2018-Expanding-Queries-for-Code-Search} use Word2vec~\footnote{\url{https://code.google.com/archive/p/word2vec/}} to extract vector representations of tokens, which is an efficient implementation of the continuous bag-of-words model (CBOW)~\cite{2013-Efficient-Estimation-of-Word-Representations-in-Vector-Space}. 
Considering that not all tokens contribute equally to the final semantic representation of code snippet, MMAN~\cite{2019-Multi-modal-Attention-for-Code-Rerieval} introduces the attention mechanism on tokens to extract the ones that are more important to the representation of a sequence of code tokens after encoding them via LSTM. The specific calculation process of LSTM and attention mechanism refer to formulas~(\ref{equ:lstm}) and (\ref{equ:attention}), respectively.
HyCoQA~\cite{2023-Hyperbolic-Code-Retrieval} transforms tokenized code snippets into numerical representations through a BERT embedding layer~\cite{2019-BERT}, because the word vectors generated by BERT can be dynamically adapted to the context provided by neighboring words.

Next, we will introduce the representation methods for the structural features of a code snippet as mentioned in Section~\ref{subsubsec:code_structural_feature_mining}.

\textbf{For the AST}, MT-CAT~\cite{2020-MP-CAT-model} converts the AST representation of a code snippet by the deterministic parser to a string using SBT~\cite{2018-Deep-Code-Comment-Generation}, and then applies FastText~\cite{2017-FastText} as the word embedding module to map a string to an embedding. TabCS~\cite{2021-TabCS} converts tree nodes into initial vector embeddings by building vocabularies. Then, since only part of the nodes can reflect the method's function, TabCS performs an attention mechanism and concatenates the weighted vectors into a feature matrix, which extracts the important nodes. The final concatenation is the feature matrix of the AST. The computation process can be referred to formulas~(\ref{equ:attention}) and (\ref{equ:attention_concatenation}), where the word sequence $\{m_1, \dots, m_n\}$ is replaced by a node sequence $\{\boldsymbol{ast}_1, \dots, \boldsymbol{ast}_n\}$. Multimodal~\cite{2021-Multimodal-Representation-for-Neural-Code-Search} transforms the AST into a novel tree structure named Simplified Semantic Tree (SST) to make the tree structure semantically better for code search, and then serializes SST to a linear token sequence by sampling tree-paths~\cite{2018-A-General-Path-based-Representation, 2021-Code-Prediction} or traversing tree-structures~\cite{2018-Deep-Code-Comment-Generation, 2018-Tree-to-Tree}. Finally, multimodal adopts a SelfAtt model to encode the tree sequence. The SelfAtt model is a transformer-based model that leverages a self-attention mechanism and BERT's positional embedding to learn from contextual information~\cite{2017-Attention-is-all-you-need, 2019-BERT}.
MMAN~\cite{2019-Multi-modal-Attention-for-Code-Rerieval} utilizes Tree-LSTM whose unit contains multiple forget gates and adopts the hidden state of the root node as the AST modality representation. Considering a node $N$ with the value $x_i$ in its one-hot encoding representation, and it has a left child $N_L$ and a right child $N_R$. The Tree-LSTM recursively computes the embedding for $N$ from the bottom up. Assume that the left child and the right child maintain the LSTM state $(\mathbf{h}_L, \mathbf{c}_L)$ and $(\mathbf{h}_R, \mathbf{c}_R)$, respectively, then the LSTM state $(h, c)$ of $N$ is computed as:
\begin{gather}
    \left.\left.(\mathbf{h}_i^{ast}, \mathbf{c}_i^{ast}) = \text{LSTM}\left(\left(\begin{bmatrix}\mathbf{h}_{iL}^{ast};\mathbf{h}_{iR}^{ast}\end{bmatrix}\right.\right., \begin{bmatrix}\mathbf{c}_{iL}^{ast};\mathbf{c}_{iR}^{ast}\end{bmatrix}\right), w(x_i)\right)
\end{gather}
where $i = 1, \dots, |x|$ and $[\cdot; \cdot]$ means the concatenation of two vectors.

\textbf{For the CFG}, as it is a directed graph, MMAN~\cite{2019-Multi-modal-Attention-for-Code-Rerieval} applies a gated graph neural network (GGNN) to represent it, which is a neural network architecture developed for graphs. 
Define a graph as $\mathcal{G}=\{\mathcal{V}, \mathcal{E}\}$, where $\mathcal{V}$ is a set of vertices $(v, \ell_v)$ and $\mathcal{E}$ is a set of edges $(v_{i}, v_{j}, \ell_{e})$. $\ell_v$ and $\ell_{e}$ are labels of vertex and edge, respectively. In the code search scenario, each vertex is the node of CFG, and each edge denotes the control flow of code. GGNN learns the graph representation through the following procedures: First, the hidden state for each vertex $v \in \mathcal{V}$ is initialized as $\mathbf{h}_{v, 0}^{cfg} = w(\ell_v)$, where $w$ is the one-hot embedding function. Then, for each round $t$, each vertex receives the vector $\mathbf{m}_{v, t+1}$, which is the ``message'' aggregated from its neighbours. It can be formulated as follows:
\begin{gather}
    m_{v, t+1} = \sum_{v^{\prime} \in \mathcal{N}(v)}\mathbf{W}_{\ell_e}\mathbf{h}_{v^{\prime}, t}
\end{gather}
where $\mathcal{N}(v)$ are the neighbours of vertex $v$. Message from each neighbour is mapped into a shared space via $\mathbf{W}_{\ell_e}$ in round $t$. GGNN updates each vertex $v$'s hidden state using the gated recurrent unit (GRU)~\cite{2014-Empirical-Evaluation-of-GRU}, which can be formulated as follows:
\begin{gather}
    \mathbf{h}_{v, t+1}^{cfg} = \mathrm{GRU}(\mathbf{h}_{v, t}^{cfg}, \mathbf{m}_{v, t+1})
\end{gather}
Finally, after $T$ rounds of iteration, the embedding representation of the CFG is obtained by aggregating the hidden state of all vertices.

\textbf{For the DFG}, given a source code $C = \{c_1, c_2, \dots, c_n\}$ with its comment $W = \{w_1, w_2, ..., w_m\}$, GraphCodeBERT~\cite{2021-GraphCodeBERT} first obtains the data flow graph. $\mathcal{G}(C) = (V, E)$ as discussed in Section~\ref{subsubsec:code_structural_feature_mining}. Next, it concatenates the comment, source code, and the set of variables $V$ in the DFG as the sequence input, and converts the sequence into an input vector by summing the corresponding token and position embeddings. Then, the model applies N transformer layers over the input vector to produce contextual representations. Specifically, it defines a graph-guided masked attention function to incorporate the graph structure into a transformer. At last, after being pre-trained on three tasks, namely masked language modeling, edge prediction, and node alignment, GraphCodeBERT can be applied to some downstream tasks, such as code search.

\textbf{For the PDG}, CRaDLe~\cite{2020-CRaDLe} first constructs a dependency matrix $\Upsilon \in \{0,1\}^{(l)\times(l)}$ according to the extracted PDG, where $l$ is the number of statements in the code snippet. The element $v_{ij} = 1$ if the $i$-th statement has a data/control dependency on the $j$-th statement; otherwise $v_{ij} = 0$. Then, it uses a one layer MLP to encode the matrix $\Upsilon$ according to the formula (\ref{equ:mlp}). Note that $\bm{\tau}_i$ in the formula (\ref{equ:mlp}) is replaced by $v_i$, and $\bm{h}_i$ is the embedding of the dependency information for each statement here. After that, CRaDLe concatenates the dependency embedding with statement-level token embedding to get the representation of the statement. Next, it adopts Bi-LSTM to encode the sequence of the statement embeddings and uses the same encoder to get the description embeddings. Finally, after calculating the cosine distance, the model will rank the code snippets and return the higher ranked ones to the programmer.

\textbf{For the program graph}, DGMS~\cite{2021-DGMS} adopts one variant of GNNs—Relational Graph Convolutional Networks (RGCNs) to learn its node embedding. Particularly, given the program graph of a code snippet $G_{e} = (\mathcal{V}_e, \mathcal{E}_e, \mathcal{R}_e)$ with nodes $e_{i}\in\mathcal{V}_{e}$ and edges $(e_i, r, e_j) \in \mathcal{E}_e$, where $r \in \mathcal{R}_e$ represents edge type, RGCN calculates the updated embedding vector $\bm{e}_{i}$ of each node $e_i \in \mathcal{V}_e$ as follows:
\begin{gather}
    \bm{e}_{i}^{(l+1)} = \mathrm{ReLU}\Bigg(W_{\Theta}^{(l)}\bm{e}_{i}^{(l)} + \sum_{r\in\mathcal{R}_{e}}\sum_{j\in\mathcal{N}_{i}^{r}}\frac{1}{|\mathcal{N}_{i}^{r}|}W_{r}^{(l)}\bm{e}_{j}^{(l)}\Bigg)
\end{gather}
where $\bm{e}_i^{(l+1)}$ is the updated embedding vector of node $e_i$ in the $(l+1)$th layer of RGCN, $\mathcal{N}_{i}^{e}$ is the set of the neighbors of node $e_i$ under the edge type $r \in \mathcal{R}_e$, $W_{\Theta}^{(l)}$ and $W_r^{(l)}$ are parameters of the RGCN model to be learned. Thus, the node embeddings $\bm{X}_{e}=\{\bm{e}_{j}\}_{j=1}^{N}\in\mathbb{R}^{(N,d)}$ for the program graph $G_{e}$ is obtained, where $d$ represents the embedding dimensions of each node.

GraphSearchNet~\cite{2023-GraphSearchNet} introduces a Bidirectional Gated Graph Neural Network (BiGGNN) to encode the program graph, which learns node embeddings from both incoming and outgoing directions for the program graph $g(V, \bm{A})$ extracted from the code snippet. During each hop $n$, for node $v$, it applies an aggregation function to take a set of incoming (or outgoing) neighboring node vectors as input and outputs a backward (or forward) aggregation vector. The summation function is selected as the aggregation function, where $N_{(v)}$ denotes the neighbors of node $v$ and $\dashv$ / $\vdash$ is the backward or forward direction.
\begin{gather}
    \bm{h}_{\mathcal{N}\dashv(v)}^n=\mathrm{SUM}(\{\bm{h}_u^{n-1},\forall u\in\mathcal{N}_{\dashv(v)}\})
    \\
    \bm{h}_{\mathcal{N}\vdash(v)}^n=\mathrm{SUM}(\{\bm{h}_u^{n-1},\forall u\in\mathcal{N}_{\vdash(v)}\})
\end{gather}
Then, the node embeddings for both directions are fused as follows:
\begin{gather}
    \bm{h}_{\mathcal{N}_{(v)}}^n=\mathrm{Fuse}(\bm{h}_{\mathcal{N}_{\lnot(v)}}^n,\bm{h}_{\mathcal{N}_{\vdash(v)}}^n)
\end{gather}
The fusion function is formulated as a gated sum of two inputs:
\begin{gather}
    \mathrm{Fuse}(\bm{a},\bm{b})=\bm{z}\odot\bm{a}+(1-\bm{z})\odot\bm{b}
    \\
    z=\sigma(\bm{W}_z[\bm{a};\bm{b};\bm{a}\odot\bm{b};\bm{a}-\bm{b}]+\bm{b}_z)
\end{gather}
where $\odot$ is the component-wise multiplication, $\sigma$ is a sigmoid function and $z$ is gating vector. Then, GRU is used to update node representations. At last, after $k$ hops of computation, the final node representation $\bm{h}_{v}^{k}$ is obtained and max-pooling is applied over all nodes $\{\bm{h}_v^k, \forall v\in V\}$ to get a $d$-dim graph representation $\bm{h}^{g}$:
\begin{gather}
    \bm{h}^g=\mathrm{maxpool}(\mathrm{FC}(\{\bm{h}_v^k,\forall v\in V\}))
\end{gather}
where $\mathrm{FC}$ is the fully-connected layer.

\finding{2}{
Similar to the query end, existing code search techniques mainly optimize the code end from two aspects: code feature mining and code feature representation. Regarding code feature mining, The usage frequency of the three code textual features (i.e., method name, API sequence, tokens) is similar. Moreover, there are many works that simultaneously utilize multiple types of code textual features. 
Among code structural features, AST is used most frequently, followed by DFG, CFG, and PDG. Some recent works have explored some new code structural features, e.g., VFG, program paragraph, and program slice. Also, some works simultaneously utilize multiple code structural features. It is worth noting that there are many code search techniques that consider both code textual features and code structural features. 
Regarding code feature representation, existing code search techniques have designed distinct code representation methods for various code features. Works in recent years have widely adopted DL-based methods to generate numerical vector representations of code textual or structural features. 
}

\section{Answering RQ3: What are the match-end optimization methods in code search studies?}
\label{sec:Answering_RQ3}
In general, developers anticipate that a code search system will prioritize the most relevant and possible code snippets as search results to facilitate their development. However, identifying the code snippet that best matches a query from thousands of snippets can be highly complex. Even with a decent understanding of the semantics of the query and the code snippet, it may still cost a large amount of time. In order to find the code snippet that matches the query efficiently and correctly, numerous methods and algorithms have been devised to optimize the match-end of the code search. 
In short, a better match-end optimization method can improve the performance of code search. It finds the most effective and efficient method to match the code snippet and query after making comprehensive use of their syntax and semantics.

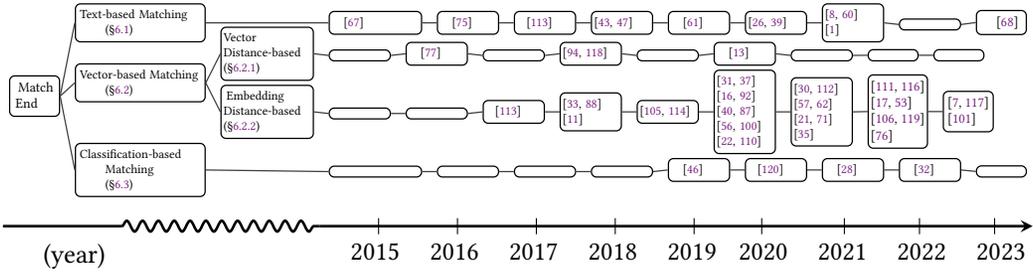
\begin{figure*}[!th]
    \centering
    \resizebox{0.98\textwidth}{!}{
        \begin{forest}
            for tree={
                grow=east,
                reversed=true,
                anchor=base west,
                parent anchor=east,
                child anchor=west,
                base=left,
                font=\large,
                rectangle,
                draw=black,
                rounded corners,
                align=left,
                minimum width=4em,
                edge+={darkgray, line width=1pt},
                s sep=3pt,
                inner xsep=2pt,
                inner ysep=3pt,
                line width=1pt,
                ver/.style={rotate=90, child anchor=north, parent anchor=south, anchor=center},
            },
            where level=1{text width=10em, font=\normalsize,}{},
            where level=2{text width=7em, font=\normalsize,}{},
            where level=3{text width=4.5em, font=\normalsize,}{},
            where level=4{text width=4.5em, font=\normalsize,}{},
            where level=5{text width=4.5em, font=\normalsize,}{},
            where level=6{text width=4.5em, font=\normalsize,}{},
            where level=7{text width=4.5em, font=\normalsize,}{},
            where level=8{text width=4.5em, font=\normalsize,}{},
            where level=9{text width=4.5em, font=\normalsize,}{},
            [\hspace{2pt}Match \\ End
                [\hspace{1.5pt}Text-based Matching \\ \hspace{22pt}(\S\ref{subsec:text_based_matching})
                    [\hspace{9.5pt}\cite{2015-CodeHow},  l=203.5pt
                        [\hspace{9.5pt}\cite{2016-Query-Expansion-Based-on-Crowd-Knowledge-for-CS}, 
                            [\hspace{7.5pt}\cite{2017-IECS}, 
                                [\cite{2018-Language-agnostic-Source-Code-Retrieval, 2018-Query-Expansion-Based-on-Code-Changes}
                                    [\hspace{9.5pt}\cite{2019-NQE}
                                        [\cite{2020-Unsupervised-Software-Repositories, 2020-recommendation-based-on-java-code-analysis-and-search}
                                            [\cite{2021-CodeMatcher, 2021-Search-for-Compatible-Code} \\
                                                \cite{2021-FACER}
                                                [
                                                    [\hspace{9.5pt}\cite{2023-Self-Supervised-Query-Reformulation}
                                                    ]
                    ]]]]]]]]
                ]
                [\hspace{1.5pt}Vector-based Matching \\ \hspace{22pt}(\S\ref{subsec:vector_based_matching})
                    [Vector \\ Distance-based \\ (\S\ref{subsubsec:vector_distance_based_matching})
                        [
                            [\hspace{9.5pt}\cite{2016-Learning-to-rank-code-examples-for-CS}
                                [
                                    [\cite{2018-Augmenting-and-Structuring-User-Query, 2018-Expanding-Queries-for-Code-Search}
                                        [
                                            [\hspace{9.5pt}\cite{2020-Enhancing-Example-based-Code-Search} 
                                                [
                                                    [
                                                        []
                        ]]]]]]]]
                    ]
                    [\hspace{2.5pt}Embedding \\ Distance-based \\ (\S\ref{subsubsec:embedding_distance-based_matching})
                        [
                            [
                                [\hspace{7.5pt}\cite{2017-IECS}
                                    [\cite{2018-Deep-Code-Search, 2018-Retrieval-on-Source-Code-A-Neural-CS} \\
                                    \cite{2018-a-neural-framework}
                                        [\cite{2019-Multi-modal-Attention-for-Code-Rerieval, 2019-Coacor}
                                            [\cite{2020-MP-CAT-model, 2020-CRaDLe} \\
                                            \cite{2020-CARLCS, 2020-CoNCRA} \\
                                            \cite{2020-csda, 2020-NJACS} \\
                                            \cite{2020-pscs, 2020-adaptive-deep-code-search} \\
                                            \cite{2020-TranS3, 2020-CodeBERT} 
                                                [\cite{2021-TabCS, 2021-Multimodal-Representation-for-Neural-Code-Search} \\
                                                \cite{2021-GraphSearchNet, 2021-DGMS} \\
                                                \cite{2021-At-CodeSM, 2021-SAN-CS} \\
                                                \cite{2021-GraphCodeBERT}
                                                    [\cite{2022-Learning-deep-semantic-model-for-CS, 2022-Incorporating-Code-Structure-and-Quality-in-Deep-Code-Search} \\
                                                    \cite{2022-fined-grained-co-attentive-representation-learning, 2022-JessCS} \\
                                                    \cite{2022-EAGCS, 2022-Enriching-Query-Semantics-for-CS-with-Reinforcement-Learning} \\
                                                    \cite{2022-SPT-Code}
                                                        [\cite{2023-CSSAM, 2023-deGraphCS} \\
                                                        \cite{2023-Hyperbolic-Code-Retrieval}
                                                        ]
                    ]]]]]]]]]
                ]
                [\hspace{1.5pt}Classification-based \\ \hspace{22pt}Matching \\ \hspace{22pt}(\S\ref{subsec:classification_based_matching})
                    [, l=203.5pt
                        [
                            [
                                [
                                    [\hspace{9.5pt}\cite{2019-ROSF}
                                        [\hspace{7.5pt}\cite{2020-OCoR}
                                            [\hspace{9.5pt}\cite{2021-Cascaded-Fast-and-Slow-Models-for-CS}
                                                [\hspace{9.5pt}\cite{2022-Accelerating-Code-Search}
                                                    []
                ]]]]]]]]]
            ]
        \end{forest}
    }
    
    \begin{tikzpicture}[x=30]
        \draw[line width=1pt] (0,0) -- (1.5,0);
        \draw[snake=snake, line width=1pt, segment amplitude=2pt, segment length=6pt] (1.5,0) -- (4,0);
        \draw[-stealth, line width=1pt,] (4,0) -- (13,0);
        
        \foreach \x in {5.0, 6.05, 7.1, 8.15, 9.2, 10.1, 11.2, 12.2, 13.2}
        \draw (\x cm,3pt) -- (\x cm,-3pt);
        
        \draw (0.9,0) node[below=3pt] {(year)} node[above=3pt] {};
        \draw (1.8,0) node[below=3pt] {} node[above=3pt] {};
        \draw (2.8,0) node[below=3pt] {} node[above=3pt] {};
        \draw (4.7,0) node[below=3pt] {2015} node[above=3pt] {};
        \draw (5.7,0) node[below=3pt] {2016} node[above=3pt] {};
        \draw (6.7,0) node[below=3pt] {2017} node[above=3pt] {};
        \draw (7.7,0) node[below=3pt] {2018} node[above=3pt] {};
        \draw (8.7,0) node[below=3pt] {2019} node[above=3pt] {};
        \draw (9.6,0) node[below=3pt] {2020} node[above=3pt] {};
        \draw (10.6,0) node[below=3pt] {2021} node[above=3pt] {};
        \draw (11.6,0) node[below=3pt] {2022} node[above=3pt] {};
        \draw (12.6,0) node[below=3pt] {2023} node[above=3pt] {};
        
    \end{tikzpicture}

    \caption{Evolution of techniques in match end}
    \label{fig:match_end_evolution}
\end{figure*}

According to the feature representation form of the query and code, existing match-end optimization techniques can be divided into three distinct groups: text-based matching, vector-based matching, and classification-based matching, as shown in Figure~\ref{fig:framework_of_code_search}. The text-based matching techniques aim to utilize the keywords from query and code snippet to make the matching, detailed in Section~\ref{subsec:text_based_matching}. 
The vector-based matching techniques are widely used in recent work. They can be further divided into two categories: vector distance-based, and embedding distance-based, detailed in Section~\ref{subsec:vector_based_matching}. 
The classification-based matching techniques aim to regard the matching tasks into classification tasks, detailed in Section~\ref{subsec:classification_based_matching}. Figure~\ref{fig:match_end_evolution} showcases the evolution of the above three types of match-end optimization techniques. In the following subsections, we will discuss them in detail and summarize their development trends.

\subsection{Text-based Matching}
\label{subsec:text_based_matching}
As a traditional matching method, text-based matching utilizes textual features such as word frequency or keywords to match code snippets with queries. These matching methods are simple and efficient and are usually adopted by the early code search methods. As shown in Figure~\ref{fig:match_end_evolution}, we have summarized a total of 12 articles that have used text-based matching methods and provided detailed summaries and overviews of them.

A basic method of text-based matching is calculating the number of common keywords shared in the text of code snippets and queries. 
CodeMatcher~\cite{2021-CodeMatcher} is a good example of directly utilizing the keywords for matching.
CodeMatcher is an IR-based code search model that inherits the advantages of DeepCS (i.e., the capability of understanding the sequential semantics in important query words).
CodeMatcher first collects metadata for query words to identify irrelevant/noisy ones. According to this preprocessed step, the collected keywords are used to launch an iterative fuzzy match on indexed method names. Then, iteratively performs the fuzzy search with important query keywords on the codebase to obtain the top $k$ candidate code snippets.
To refine the fuzzy search results, CodeMatcher designs a reranking step to measure the matching degree between query and candidate code snippets. 
During the reranking, the method name and body are regarded as two different components for the search. This is because the method name is often defined in natural language whose semantic representation is close to the query, but the method body implements the goal of the method name in programming languages. Therefore, when matching keywords for a candidate code snippet, CodeMatcher calculates the characters matched keywords in the method name of code snippets as $S_{name}$. It indicates a level of the ranked method with more overlapped tokens between the query and code snippet. 
CodeMatcher reorders code snippets based on $S_{name}$ in descending order to retrieve the final search results. In the case of tied scores, a similar approach is employed to calculate the keyword coverage scores on code methods as $S_{body}$. A higher $S_{body}$ implies better keyword matching between query and code snippets in the method body. Thus, the candidate code snippet with a higher $S_{body}$ will rank ahead.

Extended Boolean Model (EBM) is an information retrieval model that incorporates features of both the traditional Boolean model and the VSM. It combines the advantages of precisely matching from the Boolean model with the semantic similarity from VSM. 
CodeHow~\cite{2015-CodeHow} applies the EBM to consider the impact of both text similarity and APIs on code search.
According to the CodeHow introduction in Section~\ref{subsubsec:query_expansion}, it can calculate similar scores to obtain the potentially relevant APIs that match the query. Thus, CodeHow considers both the text and APIs in EBM to match the query with code snippets.
EBM conducts keyword retrieval on method name and method body of code snippets through Boolean operators and obtains similar scores.
As mentioned in CodeMatcher, the method name and body are regarded differently in matching the query and code snippets. The method name is more important than the method body. Therefore, the similarity scores for method name and method body are weighted differently in the calculation of the total score, with weights of 1.5 and 1.0, respectively. Besides, the final score also includes the API score obtained during the API understanding phase, with a weighting coefficient of 1.5. 
nIECS~\cite{2017-IECS} also utilizes EBM to conduct the similarity between the original query with code and the expanded query with code, respectively. The sum results are used to match the code snippets with the query as the final output. 

BM25 is a well-known information retrieval technique that is designed to improve the traditional TF-IDF model. BM25 is employed to assess the relevance between textual and queries, so it is commonly applied to optimize the matching end. 
Both query expansion code search models, NQE~\cite{2019-NQE} and QECK~\cite{2016-Query-Expansion-Based-on-Crowd-Knowledge-for-CS} leverage BM25 to match code snippets with queries. 
Similar to the computation for TF-IDF, it uses the following formula to give the score for the code snippets document d and a given query word $x$:
\begin{gather}
    \operatorname{BM} 25(d, x)=\operatorname{IDF}(x) \cdot \frac{\mathrm{TF}(x, d) \cdot(k+1)}{\operatorname{TF}(x, d)+k \cdot\left(1-b+b \cdot \frac{|d|}{\operatorname{avgdl}}\right)},
    \label{equ:BM25}
\end{gather}
where $TF(\cdot)$ is a function that calculates the term frequency; $IDF(\cdot)$ is a function that computes the inverse document frequency; and ${avgdl}$ is the average document length of all code snippets. $k$ and $b$ are tunable parameters that adjust the impact of term frequency on the final score, thereby enhancing the flexibility of the BM25 model. 
Based on the scores calculated using BM25, NQE and QECK output the code snippets with the highest score as the matching result for the given query. 
Furthermore, many code search methods that do not involve deep neural networks also utilize BM25 as a retrieval model to match queries with code snippets~\cite{2018-Language-agnostic-Source-Code-Retrieval,2018-Query-Expansion-Based-on-Code-Changes,2020-Unsupervised-Software-Repositories}. This indicates that BM25 is not only effective but also easy to use.

Lucene is a conventional text search engine behind many existing code search
tools. It is a versatile tool that can be embedded in various code search methods, providing efficient code retrieval capabilities. It incorporates text similarity, FQN (full quality name) of entities, and code popularity to rank the code snippets. 
The DCSE~\cite{2021-Search-for-Compatible-Code}, FACER~\cite{2021-FACER}, and SSQR~\cite{2023-Self-Supervised-Query-Reformulation} use Lucene as a retrieval tool to match the expanded queries with code snippets, demonstrating the effectiveness of their query expansion technique.

Fu et al.~\cite{2020-recommendation-based-on-java-code-analysis-and-search} regard each method as recommendation objects, denoted as documents, and store the objects in the indexes in the Lucene framework. Each document will be divided into multiple fields according to different code characteristics. When matched with the given query, they get a series of keyword tokens from the query and implement the keyword-based search in Lucene to find the retrieved results.

\subsection{Vector-based Matching}
\label{subsec:vector_based_matching}
According to Section~\ref{subsec:query_feature_representation} and Section~\ref{subsec:code_feature_representation}, it can be found that vector is the most common form of feature representations of queries and code snippets. Therefore, there are naturally more code search techniques that use vector-based matching methods to find relevant code snippets. 
Vector-based matching methods calculate distances using vectors that extract fine-grained semantic and syntactic features from code snippets and queries. 
Therefore, compared to text-based matching, these methods can retrieve code snippets that have similar features to the query, and improve overall code search performance. The second line of Figure~\ref{fig:match_end_evolution} shows that most current code search papers adopt vector-based matching methods.
Generally, there are two categories of methods for generating feature vector representations for code snippets and queries: using traditional IR methods to generate feature vectors and using DL neural networks to generate embeddings (vectors generated by DL-based methods referred to collectively as embeddings). Thereby, in this section, we divide vector-based matching into vector distance-based methods and embedding distance-based methods. Then, we will discuss and analyze vector distance matching and embedding distance matching as two separate parts.

\subsubsection{Vector Distance-based Matching}
\label{subsubsec:vector_distance_based_matching}
\noindent\newline
As mentioned earlier, vector distance-based matching methods retrieve the relevant code snippets by calculating the distance between the IR-based feature representation of both query and code snippet. 
In intuition, the closer the code snippet and the query are in these spatial distances, the more likely they are compatible in the semantics. Therefore, finding the code snippet in the vector space that is closest to the query is the core of distance-based matching.  

Before calculating the distance between code snippets and queries, representing them into the vectors is the first step of vector distance-based matching methods. 
Traditional IR methods are always used to establish the vectors. These vectors represent the semantics information of queries and code snippets.
After obtaining those vectors, they are used to match the query with code snippets in the distance and retrieve the ranking results. The most general way to measure the distance is cosine distance or Euclidean distance~\cite{1983-Boolean-Model}.
In the following, we will introduce 4 representative code search works that execute the vector distance-based matching.

As a traditional IR technique, researchers find that the Vector Space Model (VSM) is suitable for basic code search matching tasks~\cite{2016-Learning-to-rank-code-examples-for-CS, 2018-Augmenting-and-Structuring-User-Query}. VSM utilizes the common representation method TF-IDF to convert queries and code into vectors and then calculates their similarity using the cosine similarity formula:
\begin{gather}
    \cos ({v_c}, {v_q})=\frac{{v_c}^{T} {v_q}}{\|{v_c}\|\|{v_q}\|}
    \label{equ:cos_similarity}
\end{gather}
where $v_c$ and $v_q$ are the vectors of the code snippet and query, respectively. The higher the similarity, the more related the code is to the query.
Thus, we categorize the technique that employed the VSM as a part of vector distance-based matching.

Zhang et al.~\cite{2018-Expanding-Queries-for-Code-Search} also utilize VSM to calculate vector cosine distances, but in addition, they consider other features to measure the matching results between queries and code snippets. Therefore, their proposed weighted-sum ranking schema incorporates a total of five feature components in the calculation, including $fv$, which is the similarity score generated by Lucene; $fs$, which is the cosine distance similarity; $fn$, which represents keyword term frequency; $fp$, which is the number of parameters in the code example; and $fa$, which denotes the score for recommended API class names in query expansion. The scores of these components are weighted and summed to produce the final matching result.

Chen et al.~\cite{2020-Enhancing-Example-based-Code-Search} propose a semantics-based search for Java methods named Quebio. Compared with most methods in the matching end, Quebio combines a customized keyword-based search with a distance-based search to check the relevant code snippets with a given query quickly. Quebio designs the matching method in two steps. 
In the first step, a keyword-based search is employed for quick filtering, which calculates the query keyword frequencies in the code text. Only code snippets with results exceeding the threshold enter the second step. 
In the second step, the selected code snippets are preprocessed and sorted using the TF-IDF method, with the top 5 most frequently appearing words selected as the code snippet's summary. Then, VSM is used to construct vectors for these summaries, and the cosine distance between the summary vector of the code snippet and the query vector is calculated as the final retrieval result.

\subsubsection{Embedding Distance-based Matching}
\label{subsubsec:embedding_distance-based_matching}
\noindent\newline
As its name implies, embedding distance-based matching methods utilize the embeddings of the code snippet and query to calculate the distance, and optimize the ranking results. The embeddings are obtained by the deep neural network model, which can capture the semantics and structure in fine-grained, thus the ranking performance can be significantly improved.

Similar to the vector distance-based matching method mentioned above, it is necessary to measure the degree of match between queries and code snippets by computing the distance between their embeddings. 

Cosine distance is the most widely used in the field of feature similarity comparison~\cite{2017-A-Compare-Aggregate-Model-for-Matching-Text-Sequences}. Cosine distance evaluates the similarity of embeddings by calculating the cosine of their angle. 
The computation is similar to Formula~\ref{equ:cos_similarity} but replaces $v_c$ and $v_q$ with $\bm{v_c}$ and $\bm{v_q}$, representing the embeddings of code and query.
Specifically, most works~\cite{2018-Deep-Code-Search, 2018-Retrieval-on-Source-Code-A-Neural-CS, 2018-a-neural-framework,2019-Multi-modal-Attention-for-Code-Rerieval, 2020-CRaDLe, 2020-CARLCS, 2020-CoNCRA, 2020-csda, 2020-NJACS, 2020-pscs,2020-adaptive-deep-code-search, 2021-TabCS,2021-Multimodal-Representation-for-Neural-Code-Search, 2021-GraphSearchNet, 2021-DGMS, 2021-At-CodeSM, 2021-SAN-CS, 2022-Learning-deep-semantic-model-for-CS, 2022-Incorporating-Code-Structure-and-Quality-in-Deep-Code-Search, 2022-SPT-Code, 2022-fined-grained-co-attentive-representation-learning,2022-JessCS, 2022-EAGCS, 2023-CSSAM, 2023-deGraphCS} match the code snippets with query on their cosine distance in the embedding space. 

Actually, cosine similarity can be widely used not only in Euclidean spaces but also in non-Euclidean spaces (i.e., hyperbolic spaces).
As we mentioned in Section~\ref{sec:Answering_RQ1}, HyCoQA~\cite{2023-Hyperbolic-Code-Retrieval} introduces the Hyperbolic space to express connections between code snippets and their corresponding queries.
Unlike traditional Euclidean spaces, hyperbolic spaces excel in representing hierarchical structures, which frequently underlie the connection between code and its corresponding natural language description. 
Thus, after utilizing the BERT embedding layer to represent code snippets and queries into embeddings, HyCoQA utilizes a hyperbolic embedder to transform the initial embeddings into hyperbolic space.
Finally, using Formula~\ref{equ:cos_similarity} to calculate the cosine similarity between code snippet embeddings and query embeddings in hyperbolic spaces, the obtained rankings are used for the retrieved result.

Both CoaCor~\cite{2019-Coacor} and ${TranS}^3$~\cite{2020-TranS3} generate comments for code snippets to assist in code search. Therefore, they combine two cosine similarities as the final score:
\begin{gather}
    \operatorname{score}(Q, C)=\beta * \operatorname{cos}\left(e_{q}, e_{c i}\right)+(1-\beta) * \operatorname{cos}\left(e_{q}, e_{s i}\right)
    \label{equ:double_cos}
\end{gather}
where $e_{q}$, $e_{c i}$, and $e_{s i}$ are the embeddings of query, comments, and code snippets, respectively. $\beta$ is a weight parameter that ranges from 0 to 1, and $cos(\cdot)$ is the cosine similarity function.
The first cosine similarity is calculated based on the query and code snippet, and the second one is calculated based on the query and generated comments. 
While, it applies a weighting parameter to each of the two cosine distances, and then combines the results to obtain the final similarity score for ranking code snippets for a given query.

Similar to Formula~\ref{equ:double_cos}, QueCos~\cite{2022-Enriching-Query-Semantics-for-CS-with-Reinforcement-Learning} also employs a hybrid ranking approach that combines the weighted sum of two cosine similarities as the final similarity score. The difference lies in the first cosine similarity calculates between enriched queries and code snippets, and the second cosine similarity calculates between the original query and code snippets. QueCos can enrich the original query semantically through reinforcement learning, generating enriched queries. Therefore, the designed hybrid ranking approach considers both the original query and enriched queries to return the ultimate search results.

The inner product distance is also a widely used measure of the similarity between embeddings. Both CodeBERT~\cite{2020-CodeBERT} and GraphCodeBERT~\cite{2021-GraphCodeBERT} leverage the inner product distance to match the code snippets with the query. 
CodeBERT and GraphCodeBERT are the pre-trained models for the programming language, which can be used in a series of downstream tasks, including code search. In their experiments, six programming language datasets are used to fine-tune the downstream code search tasks. All the tasks calculate the inner product of code and query embeddings as relevant scores to rank candidate codes.

Euclidean distance (i.e., L2 distance) calculates the straight-line distance between two points in a multidimensional space. It is also a commonly used measure of the similarity between embeddings. MP-CAT~\cite{2020-MP-CAT-model} computes the L2 distance in the similarity module:
\begin{gather}
    \operatorname{sim}\left(e_c, e_q\right)=1-d\left(\frac{e_{c}}{\left\|e_c\right\|_{2}}, \frac{e_{q}}{\left\|e_{q}\right\|_{2}}\right)
    \label{equ:l2_distance}
\end{gather}
where $d(\cdot)$ denotes the L2 distance calculation for the dimensional code embeddings $e_c$ and query embeddings $e_q$.
The similarity module selects the code snippet with a close similarity to the given query. Finally, the output of the similarity module is regarded as the final output of the retrieved result.

\subsection{Classification-based Matching}
\label{subsec:classification_based_matching}
The classification-based matching methods are different from the distance-based method. 
Instead of calculating the distance of vectors or embeddings from the code snippets and queries or using the textual information to match the code snippets with the query, classification-based matching methods transform the matching task into a classification task.
Such methods utilize the classifier to predict the probability of semantics matching and use the predicted probabilities to rank the candidate code snippets. 
In the following, we will discuss them in detail.

To implement an effective and efficient code search system, Gotmare et al.~\cite{2021-Cascaded-Fast-and-Slow-Models-for-CS} propose a hybrid ranking framework called CASCODE, which includes the fast encoder and slow classifier to improve the performance of retrieving the search results. They prove that leveraging classification tasks involving NL-PL sequence pairs for code retrieval can achieve an optimal result. However, adopting this approach would be impractical due to the large number of candidates to be considered for each query. Therefore, they divide the retrieval process into two stages. In the first stage, the transformer encoders jointly transform the natural language query and code snippet into the embeddings and calculate the cosine distance to provide the top $k$ candidate code snippets. 
In general, this fast encoder stage ensures that the framework can quickly retrieve the top few candidate code snippets that are relatively close to the query from the code repository. 
In the slow classifier stage, they utilize a transformer encoder-based classifier to predict the top $k$ candidate code snippets from the first stage. 
The researchers' experiments demonstrate that the well-trained classifier retrieves code snippets by returning the match probabilities between queries and code snippets in semantics, resulting in excellent performance. However, using a classifier is costly and slower.
As for the limited number of candidate code snippets, a transformer classifier jointly processes the query sequence with each of the candidates to predict the probability of their semantics matching has become feasible. The predicted score of each candidate is regarded as the metric to rank the final retrieved results.

To address the problem of overlaps (e.g., ``message'' and ``msg''), which has a negative impact on retrieved results, Zhuet et al.~\cite{2020-OCoR} propose a novel neural architecture called OCoR.
In the matching end, similar to the slow classifier stage of CASCODE, OCoR also utilizes a transformer classifier to predict the probability of two classes. The first class denotes that the input natural language query and the input code are related, whereas the second class denotes that the input natural language query and the input code are unrelated. The predicted classification probability of the first class is the relevance score between the input natural language query and the code snippet.
The relevance scores are used for ranking the relevant code snippets.

To provide a better code retrieved result, Jiang et al.~\cite{2019-ROSF} propose a novel method combining both information retrieval and code classification, called ROSF. ROSF decomposes code search into two stages. Specifically, in the first stage, ROSF utilizes the IR-based method (BM25) to retrieve a candidate set that contains $N$ code snippets. This is a coarse-grained search for identifying a few relevant candidate code snippets. In the next fine-grained re-ranking stage, ROSF views the problem of ranking the candidate code snippets as a multi-class classification task. For each instance of the candidate set for a new query, ROSF employs the learned linear predictor function to predict the probabilities of four possible relevance scores. In other words, each instance has four probability values corresponding to four relevance scores. Then, the relevance score with the maximum probability value is selected as the predicted relevance score for the instance. Among the candidate sets, ROSF finally sorts the subset containing the code snippets with predicted $score 4$ according to the predicted probability values in descending order.

To accelerate the retrieval efficiency of deep learning-based code search approaches, Gu et al.~\cite{2022-Accelerating-Code-Search} propose a novel approach called CoSHC, which adopts the recall and re-rank mechanism with the integration of code classification and deep hashing to improve code search performance. CoSHC first generates the code and description embeddings from deep learning networks. Then, a deep hashing module is utilized to generate the corresponding binary hash codes for the embeddings in binary hashing space. Since the capacity of binary hashing space is very limited compared to Euclidean space, they cluster the source code whose representation vectors are close to each other into the same category. 
After obtaining the code representation categories in hashing space, the classifier in the category prediction module will calculate the probability distribution of categories for the given query. The number of code candidates $R_i$ for each category $i$ will be recalled according to this probability distribution, which can be computed as: 
\begin{gather}
    R_{i}=\min \left(\left\lfloor p_{i} \cdot(N-k)\right\rfloor, 1\right), i \in 1, \ldots, k,
\label{equ:CoSHC}
\end{gather}
where $p_{i}$ is the predicted probability for category $i$; $N$ is the total recall number of source code and $k$ is the number of categories.
In the final re-ranking stage, the original representation vectors of these recalled code candidates will be retrieved and utilized for the cosine similarity calculation.

\finding{3}{
There are three optimization methods utilized in the match end by existing code search techniques, i.e., text matching-based methods, vector matching-based methods, and classification-based methods. All three types of methods have received widespread attention. Among them, with the popularity of DL technology, vector matching-based methods (especially embedding distance-based methods) are the most common. Although classification-based methods are also based on embeddings, they require training additional classification layers, which means they have higher training costs. It is foreseeable that embedding distance-based methods will remain mainstream for some time in the future.}

\section{Challenges and Opportunities for Future Work}
\label{sec:outlook_and_challenges}
After analyzing the existing code search techniques in both advantages and disadvantages, we can suggest that further research will also have practical and research significance. From the practical significance of implementing code search, these excellent techniques will certainly help us solve the difficulties in the development of software. On the other hand, code search still has many challenges to be solved. We pose some of the challenges and opportunities that have emerged from the papers below.

\subsection{Challenges}
\label{subsec:challenges}
\textbf{Challenges in Query Feature Mining.} Query reduction aims to remove noisy/redundant terms of the query and preserve important query features. However, it may remove certain terms might result in a loss of semantic context, making it challenging to capture the user's intent accurately. In addition, users might have different expectations regarding the reduction process, and striking a balance between code search precision and recall is challenging. 
Query expansion aims to add new terms to the original query to enrich query features. However, the relevance and quality of the newly added terms play a significant role. Using irrelevant or inaccurate terms can hinder the search process. How to control the quality of extensions is challenging. 
Query transformation aims to transform the original query feature into an alternative form feature, e.g., code API. However, transforming a query into an appropriate code-related feature while preserving its semantic meaning is challenging. Ensuring accurate transformation is crucial for retrieving relevant results. Additionally, some users search for code in domain-specific programming languages that might not have direct mappings from natural language queries. Adapting query transformation rules to handle domain-specific code retrieval is a challenge.

\noindent\textbf{Challenges in Query Feature Representation.} IR-based query feature representation methods represent queries as some form of index (such as keyword terms and vectors). However, queries and code snippets might use different vocabularies or terminologies, leading to a mismatch in the representation. How to align the vocabularies effectively is crucial and challenging for accurate retrieval. 
DL-based query feature representation methods apply deep neural networks to encode the given query feature to produce embeddings. However, queries often contain ambiguous terms or phrases that can have multiple meanings in the context of code. It is challenging for deep learning models to discern the intended meaning of ambiguous queries accurately. In addition, queries can vary significantly in length, making it challenging to process them uniformly. The applied models need to handle variable-length inputs effectively.

\noindent\textbf{Challenges in Code Feature Mining.} Although existing code search techniques have mined various code features to represent code semantics, how to fully utilize each feature is still full of challenges. In addition, there is information overlap between different code features. For example, Token and the labels of AST leaf nodes overlap. Therefore, how to deal with this redundant information is also a challenge when wanting to consider multiple features simultaneously. Finally, whether structural features without Token information (e.g., syntactic structure in AST, control flow in CFG, and data flow in DFG) actually help facilitate code representation and code search tasks still lacks systematic research. The diversity of feature parsing tools and preprocessing methods makes such systematic research challenging.

\noindent\textbf{Challenges in Code Feature Representation.} Using deep learning techniques to represent code features has become mainstream nowadays. Compared to traditional IR-based code feature representation methods, the code representations generated by DL-based cod feature representation methods can better capture the semantics of code. However, DL-based methods require large, high-quality labeled datasets for effective training. Obtaining such datasets with accurately labeled code snippets can be challenging, especially for specific programming languages or domains. 
In addition, code snippets also vary significantly in length, from short expressions to lengthy functions. Designing representations that handle variable-length inputs effectively is a challenge. 
Moreover, as mentioned earlier, code contains multiple textual and structural features. Integrating these diverse features into a cohesive representation without losing essential information is challenging. 

\noindent\textbf{Challenges in Query-Code Matching.} Since different developers have different wording habits, this will cause the text-based matching method to fail to retrieve the expected results. Therefore, mapping relationships between query vocabulary and code vocabulary is challenging for methods of this kind. For vector distance-based matching methods, it is necessary for the query vectors and code vectors to be in a unified vector space, and the distance between vectors should accurately reflect semantic relevance. However, whether utilizing IR techniques or DL techniques, mapping queries and code snippets to a unified vector space while preserving their respective semantics is a challenging task. 
Instead of directly calculating distance using query and code vectors, classification-based matching methods involve concatenating the two vectors and utilizing a classifier for binary classification. However, correctly concatenating query and code vectors is challenging because vectors are abstract, and deep learning is low in interpretability. 
Last but not least, an aspect that requires more attention is the efficiency challenge faced by all query-code matching methods when dealing with extensive code search corpus. Therefore, designing efficient matching methods is also a challenging task.

\subsection{Opportunities}
\label{subsec:opportunities}

\textbf{Opportunities in Query End:} As discussed in Section~\ref{sec:Answering_RQ1}, the effectiveness of query reduction, expansion, and transformation methods has been validated in multiple code search techniques, demonstrating their ability to enhance accurate code retrieval. Therefore, the simultaneous integration of these three methods presents an opportunity to enhance query feature mining and subsequently improve code search performance. 
Whether it is possible to utilize open-source data to establish alignment between query terms and commonly used tokens in code before inputting them into the model, thereby enhancing the representation of query features, is also a good opportunity. 

\noindent\textbf{Opportunities in Code End:} While existing research has indicated that combining multiple code features can enhance code search performance, it appears that each code feature has not yet exerted its maximum value. Therefore, there are still ample opportunities for research in the integration of multiple code features (especially, multimodal code feature data). 
Additionally, a thorough analysis of the impact of various code features on code representation and subsequent code search performance is a valuable research opportunity. 
Lastly, the recent emergence of large language models (e.g., ChatGPT~\footnote{\url{https://chat.openai.com}}), particularly large code models (e.g., StarCoder~\footnote{\url{https://github.com/bigcode-project/starcoder}}), has provided new opportunities for code feature exploration and representation. For instance, the code feature extraction and representation capabilities of large language models are worth investigating.

\noindent\textbf{Opportunities in Match End:} First, regarding the challenges mentioned in Section~\ref{subsec:challenges}, a research opportunity is investigating how to map queries and code snippets to a unified vector space while preserving their respective semantics. 
Second, the impact of the concatenation of query and code vectors on classification-based query code matching needs to be explored, which also poses research opportunities. 
Last, there are still numerous research opportunities to enhance the efficiency of matching queries and code. For example, one could explore first clustering a large number of code snippets and then performing query code matching.

\section{Threats to validity}
\label{sec:threats_to_validity}
In this chapter, we explore the validity threats to our study. Validity concerns the relationship between the research results and the actual situation, as well as how the conclusions might be incorrect. We have categorized the threats to the validity of our study into two types, including internal validity and external validity. Our discussions on these identified threats are as follows:

\subsection{Threats to Internal Validity}
\label{subsec:Threats_to_Internal_Validity}
The threats to internal validity refer to experiment errors and human biases~\cite{2014-Automated-Construction-of-Database}. 
In our study, we employ a cautious research strategy to mitigate threats to internal validity. 
Firstly, we establish three specific research questions covering the different essential dimensions of the code search technique (detailed in Section~\ref{subsec:research_questions}). Subsequently, we utilize the PIO approach to generate keywords from these research questions and collect papers from six widely used electronic databases (including DBLP, Google Scholar, IEEE Explore, etc.). These collected papers under multiple rounds of manual filtering and screening based on our defined research scope and criteria (detailed in Section~\ref{subsec:search_strategy},~\ref{subsec:study_selection}). From the remaining 1427 papers, we meticulously review 68 of them, analyzing their technical aspects according to the three research perspectives. Additionally, we explore and discuss potential future developments based on existing work within each perspective. 
As a result, the systematic exploration has instilled confidence in our work and established a certain level of effectiveness in the research outcomes.

\subsection{Threats to External Validity}
\label{subsec:Threats_to_External_Validity}
External validity concerns the extent to which the results can be generalized beyond the scope of the study, even when specific cause-and-effect relationships have been established in the study. Threats to external validity involve the generalizability of reported research findings~\cite{2014-Automated-Construction-of-Database}.
For our study, we meticulously select 68 papers from six commonly used electronic databases. Hence, these papers likely cover the core studies in the code search field. Therefore, the conclusions and findings drawn in our study can potentially apply to the techniques inadvertently missed in this paper or those emerging as new techniques in the near future. Furthermore, within each research question section, we have conducted theoretical qualitative analyses. Hence, we believe that external validity threats are not particularly challenging for this study.

\section{Conclusion}
\label{sec:conclusion}
This article provides a 3-dimensional survey of the technological developments in code search over the past thirty years. 
This survey focuses on the three core components of code search technology, i.e., query understanding component, code understanding component, and query-code matching component. 
We classify and discuss the optimization techniques proposed for each component. 
Specifically, we divide the techniques for optimizing the query and code ends into two parts, namely feature mining and feature representation, respectively. For the match end, we categorize existing optimization techniques into three major classes, i.e., text-based matching, vector distance-based matching, and classification-based matching. For each end (each optimization category), we provide a detailed introduction to some representative optimization technologies and summarize the technology development trends. 
Based on the comprehensive observation of optimization techniques proposed in existing code search papers, we summarize some challenges that still need to be addressed and suggest research opportunities.

\section{Acknowledgments}
This work is supported partially by the National Natural Science Foundation of China (61932012, 62141215), and the Program B for Outstanding PhD Candidate of Nanjing University (202201B054).

\bibliographystyle{ACM-Reference-Format}
\bibliography{reference}





\end{document}